\documentclass[prb,aps,superscriptaddress,reprint]{revtex4-1}

\usepackage{graphicx}% Include figure files
\usepackage{dcolumn}% Align table columns on decimal point
\usepackage{bm}% bold math
\usepackage{setspace}
\usepackage{amsmath}
\usepackage{epsfig}
\usepackage{todonotes}
\usepackage[normalem]{ulem}

\newcommand{\fg}{f_{\mathrm{G}}}

\newcommand{\ms}{M_{\mathrm{S}}}

\newcommand{\hs}{H_{\mathrm{S}}}

\newcommand{\ke}{\kappa_\mathrm{E}}
\newcommand{\kz}{\kappa_\mathrm{Z,OOP}}
\newcommand{\kdip}{\kappa_\mathrm{D,IP}}
\newcommand{\kdoop}{\kappa_\mathrm{D,OOP}}

\begin{document}

\title{Exchange-mediated, non-linear, out-of-plane magnetic field dependence of the ferromagnetic vortex gyrotropic mode frequency driven by core deformation}

\author{Jasper P.~Fried}
\email{21114359@student.uwa.edu.au}
\affiliation{School of Physics, M013, University of Western Australia, 35 Stirling Hwy, Crawley WA 6009, Australia.}
\author{Hans~Fangohr}
\affiliation{Engineering and the Environment, University of Southampton, Southampton, United Kingdom}
\author{Mikhail~Kostylev}
\affiliation{School of Physics, M013, University of Western Australia, 35 Stirling Hwy, Crawley WA 6009, Australia.}
\author{Peter J.~Metaxas}
\email{peter.metaxas@uwa.edu.au}
\affiliation{School of Physics, M013, University of Western Australia, 35 Stirling Hwy, Crawley WA 6009, Australia.}

\date{\today}% It is always \today, today,

%75.70.Kw, 	vortices in magnetic thin films,
%75.78.Cd 	umag sims of dynamics
% 76.50.+g 	Ferromagnetic, antiferromagnetic, and ferrimagnetic resonances; spin-wave resonance

%150 words
\begin{abstract}
We have performed micromagnetic simulations of the vortex gyrotropic mode resonance in a range of disk geometries subject to spatially uniform out-of-plane magnetic fields. For disks of small lateral dimensions, we observe a drop-off in the mode's frequency for field amplitudes approaching the disk saturation field. This non-linear frequency response is shown to be associated with an increased vortex core deformation, which results from the demagnetizing field created when the core is shifted laterally. Such deformation results in an increase in the average out-of-plane magnetization of the displaced vortex state, which through an exchange contribution, leads to a sharp decrease in the vortex stiffness coefficient. It is this decrease in the vortex stiffness coefficient which leads to the non-linear field dependence of the gyrotropic mode frequency. 
\end{abstract}

\maketitle

\section{Introduction}

Magnetic vortices are ground state magnetization configurations that form spontaneously in magnetic elements such as (sub)micron disks \cite{Ha2003,Guslienko2004}. They consist of a curling magnetization that turns out-of-plane at the disk center over a nano-scale region known as the vortex core \cite{Cowburn1999-2,Shinjo2000,Wachowiak2002}. The vortex state has been extensively studied in the past decade, partly due to its rich excitation spectrum \cite{Ding2014,Kammerer2011,Aliev2009,Boust2004,Ivanov2005}. The lowest frequency excitation of a magnetic vortex is the gyrotropic mode, which corresponds to orbit-like motion of the vortex core about its equilibrium position \cite{Choe2004,Guslienko2002,Park2003}. This mode can be excited by the application of a time varying in-plane magnetic field, and its frequency, $\fg$, depends strongly on the vortex magnetization configuration \cite{Guslienko2002}. It is therefore not surprising that the application of an external magnetic field, which modifies the vortex spin structure through the addition of a Zeeman term, will alter $\fg$ \cite{deLoubens2009,Yoo2011,Buchanan2006,Yakata2013,Ivanov2002}. Indeed this field tune-ability has been exploited for several novel applications such as tunable electronic oscillators for radiofrequency signal generation \cite{Dussaux2010,Lebrun2014,Pribiag2007} and frequency based \cite{Braganca2010} magnetic field sensing \cite{Fried2016}. 

All of the above applications that make use of the field tune-ability of $\fg$ rely on a detailed understanding of how the gyrotropic frequency changes with field amplitude. It has previously been demonstrated that low amplitude, spatially uniform magnetic fields perpendicular to the disk plane modify the vortex magnetization configuration so as to result in a linear change in $\fg$ with increasing field amplitude \cite{deLoubens2009,Yoo2011,Pribiag2007}. However, there can be a significant deviation from this linear frequency behavior when the applied field amplitude is close to the disk saturation field \cite{Dussaux2010,DussauxThesis}. In this work we carry out a detailed study of the phenomena underlying this non-linear field dependence. Such an analysis will no doubt be important for the further development of the many proposed technologies which exploit the field dependence of the gyrotropic frequency. 

This paper is set out as follows. In Sec.~\ref{smethod} we give a brief description of the micromagnetic simulation technique. In Sec.~\ref{snonlinear} we present simulation results exhibiting a drop-off in the gyrotropic frequency for small disks in large out-of-plane fields. In Sec.~\ref{sdeformed} we look at the effect of an external out-of-plane field on vortex core deformation during gyrotropic motion. Here it is shown that in the case of a small disk in large out-of-plane fields, there is an increased deformation of the core magnetization profile due to the out-of-plane demagnetizing field created when the core is shifted laterally. Finally in Sec.~\ref{scalcs}, we relate this increased core deformation to the non-linear field dependence of the vortex gyrotropic resonance frequency by performing a detailed analysis of the field dependence of the vortex stiffness coefficient and gyroconstant. Here it is shown that for small disks in large out-of-plane fields, deformation of the vortex core driven by the demagnetizing field leads to a sharply decreasing exchange contribution to the vortex stiffness coefficient, resulting in a drop-off in the gyrotropic frequency. 

\section{Micromagnetic Simulation}
\label{smethod}

We will concentrate primarily on simulations run in MuMax3 \cite{Vansteenkiste2014} for Permalloy-like disks with saturation magnetization $\ms=800\,$kA/m, exchange constant $A=$ 13$\,$pJ/m, magnetic damping parameter $\alpha=0.008$ and nil intrinsic anisotropy. A cell size of $3\times3\times3.75\,$nm$^{3}$ was used for the range of simulated disk geometries. For simplicity, in this work we will only consider the case of a vortex with a positive core polarity which is aligned with the external out-of-plane field.  The equilibrium magnetization configuration was first found by initializing a vortex like spin structure, applying the desired static external field and evolving the magnetization (without precession) to reach the minimum energy state. To induce gyrotropic motion (and hence find $\fg$), vortex core dynamics were driven via the application of a transverse sinc pulse field: $A\sin(\omega(t-t_0))/(\omega(t-t_0))$. This excites modes up to a cut-off frequency of $\omega/2\pi=30\,$GHz. The resonant eigenfrequencies can then be determined by performing a Fourier analysis of the time dependent spatially averaged in-plane magnetization \cite{Baker2016}. We note that while this process also excites higher frequency spin waves \cite{Aliev2009,Park2005,Ivanov2005,Buess2004}, this work will focus on the lowest frequency excitation corresponding to the vortex gyrotropic mode. 

\section{Non-linear frequency responses}
\label{snonlinear}

\begin{figure}[t]
	\centering
	\includegraphics[width=7.5cm]{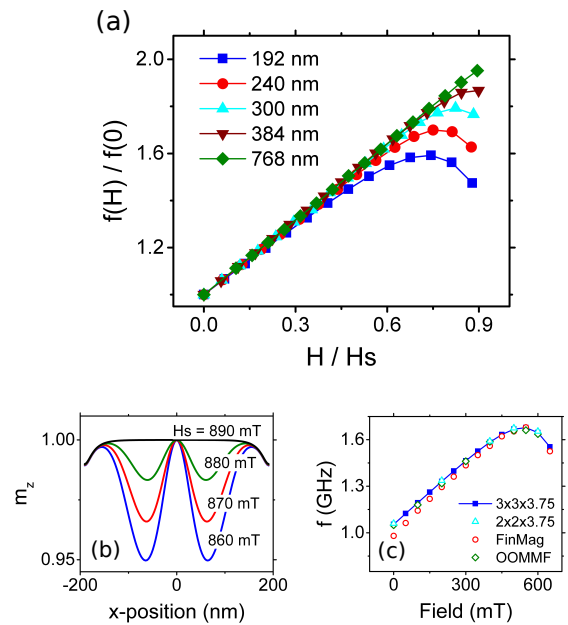}
	\caption{(a) Simulated gyrotropic frequency as a function of uniform out-of-plane field amplitude for disks of thickness $L=30\,$nm and a range of diameters. $\fg$ has been normalized by the simulated frequency in zero out-of-plane field while the field amplitude has been normalized by the disk saturation field. (b) Out-of-plane magnetization for a slice through the disk center for a 384$\,$nm diameter disk in increasing field amplitudes (c) Comparison of the simulated values of $\fg$ found using  MuMax3 (for two different cell sizes) with the frequencies found using OOMMF and FinMag (disk diameter is 192$\,$nm).}
	\label{fdropoff}
\end{figure}

The simulated gyrotropic frequencies as a function of uniform out-of-plane field amplitude, are shown in Fig.~\ref{fdropoff}(a) for several disk diameters with thickness $L=30\,$nm. To be able to obtain comparable results across all disk geometries, $\fg$ has been normalized by the simulated frequency found in zero out-of-plane field for each disk size. Likewise the field amplitude has been normalized by the disk saturation field, $H_\mathrm{S}$, which was found by stepping through fields of increasing amplitude and plotting the equilibrium out-of-plane magnetization to determine when the vortex state is no longer present. The result of such a process is shown in Fig.~\ref{fdropoff}(b) for the 384 nm disk diameter where the saturation field is found to be $885\pm5\,$mT.

We see that while the gyrotropic frequency scales linearly in low field amplitudes, there is a drop-off (or mode `softening') in $\fg$ when the applied field is close to $\hs$. Moreover, this drop-off becomes less distinct as the disk radius is increased, with $\fg$ being almost linear up to the saturation field for the 768$\,$nm disk. This drop-off in $\fg$ has been previously observed via micromagnetic simulation and experiment \cite{Dussaux2010,Dussaux2011} in disks with a diameter of 170$\,$nm.

We finally note that we have confirmed the frequency behavior in Fig.~\ref{fdropoff}(a) for the 192 nm disk when decreasing the mesh size to $2\times 2 \times 3.75\,$nm$^3$. We have also compared these frequencies to results obtained using a similar time domain method in OOMMF\cite{oommf}  ($3\times3\times3\,$nm$^{3}$ cell size) and an eigenmode method \cite{dAquino2009,Metaxas2014} in FinMag (derived from Nmag\cite{Fischbacher2007}) with good agreement being found across the range of applied fields [Fig.~\ref{fdropoff}(c)]. We used a  non-uniform finite element mesh in FinMag with a characteristic internode length of 3$\,$nm  at the disk center which smoothly transitioned to  5$\,$nm at the disk edge.

\section{Core dynamics and deformation}
\label{sdeformed}

To begin to understand the cause of the non-linear frequency behavior, it is necessary to look at how the vortex magnetization configuration changes \emph{during} gyrotropic motion. While this might initially seem unrelated to the observed drop-off in $\fg$, in later sections it will be shown that deformation of the vortex core magnetization profile during gyrotropic motion is the primary cause of the frequency drop-off. 

\subsection{Differentiating static and dynamic core deformation}

\begin{figure}[t!]
	\centering
	\includegraphics[width=7.5cm]{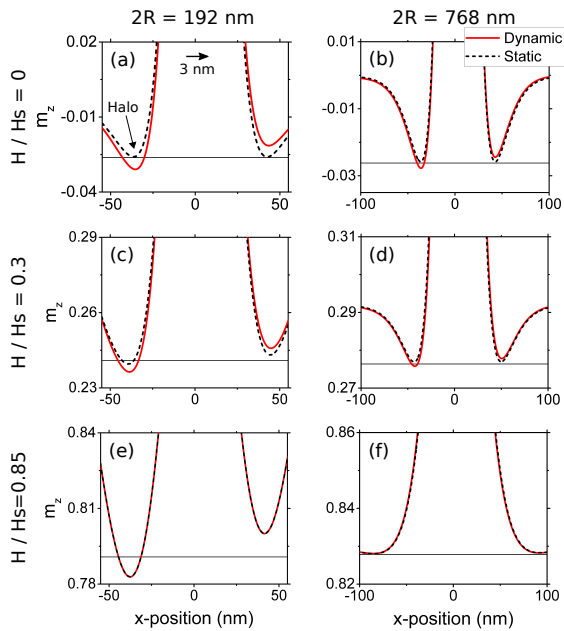}
	\caption{Out-of-plane magnetization profile close to the vortex core for a 192 and 768$\,$nm disk diameter in several normalized field amplitudes. The solid red lines show the magnetization profile while the core is undergoing gyrotropic motion. The black dashed lines show the magnetization profile when the core has been shifted to a new equilibrium position by a static in-plane magnetic field. The horizontal black lines reference the minimum of the magnetostatic halo when the vortex core is stationary and at the disk center.}
	\label{fdyncore}
\end{figure}

Figure \ref{fdyncore} shows slices of the out-of-plane magnetization near the vortex core for a range of normalized field amplitudes. These one-dimensional slices of the thickness averaged magnetization have been taken so as to intersect the vortex core center and the disk center. For simplicity we will concentrate on results obtained for the 192 and 768$\,$nm disks.  The solid red lines are for  gyrotropically resonating (`dynamic') vortex cores whereas the dashed black lines are for `static' vortex cores which have been shifted to new equilibrium positions by static in-plane magnetic fields. All core displacements in Fig.~\ref{fdyncore} are $\approx$ 3$\,$nm in the positive $x$ direction to enable comparison between the different profiles.

At zero out-of-plane field, the static displaced cores are highly symmetric in both the small and large disk. While this symmetry is retained for the larger disk size when the out-of-plane field is increased, for the 192$\,$nm disk the static displaced cores become deformed with the minimum of the magnetostatic halo\cite{Gaididei2010} [labelled in Fig.~\ref{fdyncore}(a)] moving up from its position for a centered vortex (referenced by the solid horizontal lines) in the direction of core displacement. This is accompanied by a downward shift of the magnetostatic halo on the side of the core which is away from the direction of core displacement. Notably the magnitude of this asymmetry in the magnetization profile of statically displaced cores in the 192$\,$nm disk increases as the external field amplitude is increased.
 
Dynamically displaced cores are however clearly asymmetric for both disk sizes in zero out-of-plane field. As the out-of-plane field is increased, the asymmetry of the statically and dynamically displaced core profiles become comparable. Indeed, for $H/\hs$ = 0.85, the two profiles are almost identical for both disks (i.e.~highly asymmetric in the 192$\,$nm disk and highly symmetric in the 768$\,$nm disk). 

In the following section, this behavior will be shown to be due to two forms of core deformation. Namely for a disk in zero or low out-of-plane fields core deformation is predominantly due to the gyroforce resulting from vortex core motion and is thus only seen in the dynamically displaced profiles. However for large out-of-plane fields asymmetry in the core magnetization profile is driven by the demagnetizing field created when the core is shifted laterally and is thus observed in both statically and dynamically displaced vortex cores.

\subsection{Gyrofield driven core deformation}

Deformation of the magnetization profile of a dynamic vortex core (i.e.~one undergoing gyrotropic motion) has previously been observed in the absence of an out-of-plane field \cite{Yamada2007,Waeyenberge2006,Novosad2005} and was attributed to the gyroforce \cite{Guslienko2008b} resulting from the motion of the non-uniform magnetization configuration. The effect of the gyroforce can be described by an effective field, the gyrofield, the out-of-plane component of which causes core deformation and can be calculated by \cite{Guslienko2008b}:
\begin{equation}
h_z(r,t)=\frac{1}{\gamma}\frac{(\mathbf{m}\times\mathbf{\dot{m}})_z}{(m_z+p)^2}
\label{egfield}
\end{equation}
where $\gamma$ is the gyromagnetic ratio, $\mathbf{m}$ is the magnetization unit vector, $m_z$ is the out-of-plane magnetization, $p$ is the vortex core polarity and $\mathbf{\dot{m}}=-(\mathbf{\dot{X}}\cdot\mathbf{\nabla})\mathbf{m}$ where $\mathbf{\dot{X}}$ is the core velocity which can be determined from the vortex core position and the gyrotropic frequency.

\begin{figure}[t!]
	\centering
	\includegraphics[width=8.3cm]{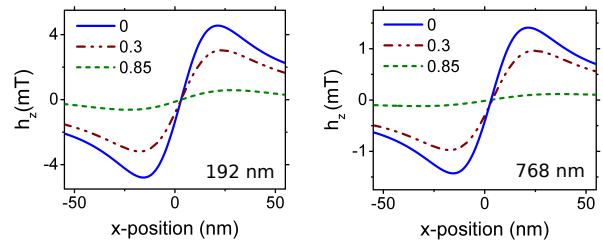}
	\caption{The gyrofield close to the vortex core as calculated using Eq.~(\ref{egfield}) for disk diameters of 192 and 768$\,$nm and three different normalized out-of-plane field values.}
	\label{fgyrofield}
\end{figure}

In Fig.~\ref{fgyrofield} we have numerically calculated the gyrofield for each core profile shown in Fig.~\ref{fdyncore}. For both disk diameters the gyrofield decreases as the out-of-plane field amplitude is increased. This is because increasing the field leads to a higher degree of out-of-plane canting of the curling magnetization which reduces the spatial magnetization gradient  and thus the gyrofield (via $\mathbf{\dot{m}}$ which depends on $\nabla \mathbf{m}$). The gyrofield profile is consistent with the core asymmetry of dynamically displaced cores observed in Fig.~\ref{fdyncore}, i.e.~in the direction of core displacement the halo is shifted upward due to a positive gyrofield. On top of this, the weakening of the gyrofield with increasing out-of-plane fields explains why, as observed in Fig.~\ref{fdyncore}, the static and dynamic displaced core profiles become similar as the field amplitude is increased (due to the dynamic source of deformation, the gyrofield, being strongly reduced).  

\begin{figure}[t!]
	\centering
	\includegraphics[width=7.5cm]{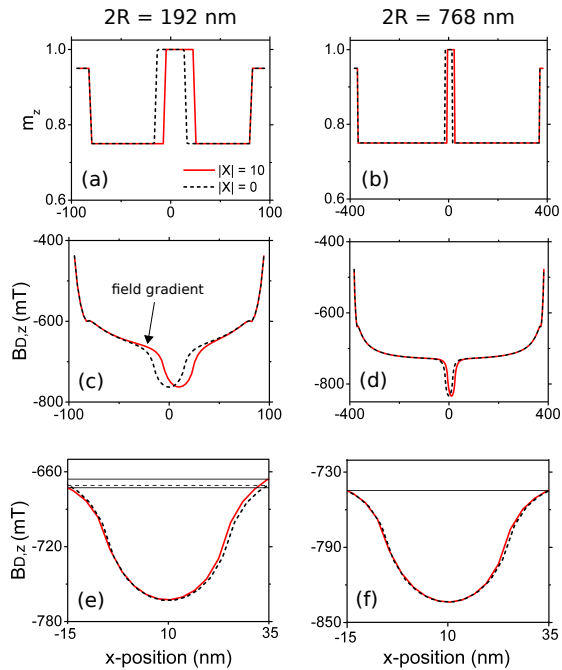}
	\caption{(a,b) Slices through the created vortex-like configuration for the (a) 192 and (b) 768$\,$nm disk diameter. The solid red lines show the magnetization profile for a core shifted 10$\,$nm in the positive $x$ direction. The black dashed lines show the magnetization profile for a centered vortex core. (c,d) Demagnetizing field profiles for a displaced and centered vortex core in the (c) 192 and (d) 768$\,$nm disk. (e,f) Demagnetizing field close to the vortex core for the (e) 192 and (f) 768$\,$nm disk. The solid horizontal lines reference the magnetostatic field 25$\,$nm either side of the core maximum for the case of a displaced vortex core. The dashed horizontal line marks the magnetostatic field 25$\,$nm either side of the core maximum for the case of a centered vortex core.}
	\label{fmademag}
\end{figure}

\subsection{Demagnetizing field driven core deformation}

The deformation of dynamic and static displaced vortex cores for small disks in high out-of-plane fields can be shown to be a magnetostatic effect by looking at the demagnetizing field created by statically shifted, \emph{non-deformed} vortex cores. Note that it is necessary to look at the demagnetizing field created by symmetric, non-deformed core profiles to avoid confusion related to whether asymmetries in the demagnetizing field are the cause of the core deformation or a consequence of it. To do this we created simple magnetization configurations crudely approximating that of a shifted (by 10$\,$nm) and unshifted vortex in the presence of a high out-of-plane field [Fig.~\ref{fmademag}(a) and (b)]. The demagnetizing field profile (averaged across the disk thickness) created by these `artificial' magnetization configurations is shown Fig.~\ref{fmademag}(c) and (d).

As shown in Fig.~\ref{fmademag}(c) and (d), for both disk sizes the $z$ component of the magnetostatic field is strongest around the vortex core where the out-of-plane magnetization is highest. The field amplitude then drops-off close to the disk lateral boundary since here there is no adjacent moments to reinforce the demagnetizing field. For the smaller disk size, the reduced lateral dimensions mean that these edge effects result in a spatial gradient of the demagnetizing field close to the disk center [Fig.~\ref{fmademag}(c)]. This field gradient close to the vortex core in the smaller disk size leads to an asymmetry in the demagnetizing field created when the core is shifted laterally. Namely, in the direction of core displacement, the demagnetizing field a given distance from the core center will decrease in amplitude (become less negative) as it will be closer to the disk edge. Similarly, on the side away from the direction of core displacement the magnetostatic field will increase in amplitude (become more negative) as it will be farther away from the disk's lateral boundary. We would expect this effect to be significantly reduced for the larger disk diameter as here the influence of the disk edge on the created demagnetizing field becomes negligible close to the vortex core.

Such asymmetry in the demagnetizing field profile is seen in Fig.~\ref{fmademag}(e) which shows the same field profile as Fig.~\ref{fmademag}(c), however here we concentrate on the field 25$\,$nm either side of the core maximum. Note that the field profile of the non-displaced vortex has been translated 10$\,$nm in the positive $x$ direction to enable comparison between the two cases. For the smaller disk size there is a clear difference between the field amplitude 25$\,$nm either side of the core maximum (referenced by the solid horizontal lines) for the vortex with a displaced core. As expected this asymmetry is significantly reduced for the larger disk size [Fig.~\ref{fmademag}(f)]. Notably the asymmetry in the demagnetizing field profile is consistent with deformation of statically shifted vortex cores observed in Fig.~\ref{fdyncore}. Namely in the direction of core displacement the height of the magnetostatic halo is increased consistent with a reduced negative demagnetizing field.

We also note that when compared to the demagnetizing field created by a vortex with a centered core, the field increase to the right of the core is greater than the decrease to the left. This can be seen in Fig.~\ref{fmademag}(e) when comparing the distance between the solid and dashed horizontal lines: the dashed line references the magnetostatic field 25$\,$nm either side of the core for a centered vortex profile. The change in the demagnetizing field's `net' direction when the core is displaced is therefore aligned with the core polarity, which is in contrast to gyrofield driven deformation where the field opposing the core is always greater than that reinforcing it \cite{Guslienko2008b,Yamada2007,Waeyenberge2006}. Such behavior in the gyrofield profile can be identified in Fig.~\ref{fgyrofield} where, for the smaller disk in zero out-of-plane field, the maximum gyrofield opposing the vortex core is $\approx$ -4.80$\,$mT compared to $\approx$ 4.55$\,$mT reinforcing the core polarity. 

\begin{figure}[t!]
	\centering
	\includegraphics[width=6cm]{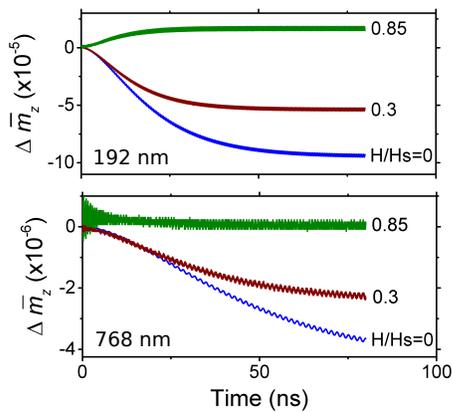}
	\caption{Plots of the evolution of the spatially average out-of-plane magnetization as a function of time since gyrotropic motion was induced for a 192 and 768$\,$nm disk in various out-of-plane fields. The gyrotropic motion has been driven by a in-plane sinusoidal field with frequency equal to the $\fg$ values in Fig.~\ref{fdropoff} found via the sinc pulse excitation detailed in Sec.~II.}
	\label{fdmz}
\end{figure}

\subsection{Time evolution of the out-of-plane magnetization}

The differing net directions of the field produced by the demagnetizing field and gyrofield driven core deformation results in different changes in the spatially averaged out-of-plane magnetization component ($\bar{m}_z$) when the vortex core is dynamically displaced. This effect is seen in Fig.~\ref{fdmz} where we have plotted $\bar{m}_z$ as a function of time after the start of gyrotropic motion for the two disk sizes in a range of field amplitudes. For both diameters, when in zero out-of-plane field, $\bar{m}_z$ decreases as the core oscillation amplitude increases with time. This is consistent with vortex core deformation at low out-of-plane fields being driven by the gyrofield which creates a net field that opposes the core polarity. For the larger disk size this effect decreases with increasing out-of-plane field since the amplitude of the gyrofield decreases (Fig.~\ref{fgyrofield}) and demagnetizing field driven deformation is minimal. Indeed at high out-of-plane fields the gyrofield has decreased to an extent where there is almost no vortex core deformation [Fig.~\ref{fdyncore}(f)] resulting in the change in $\bar{m}_z$ being $\approx 0$. 

For the smaller disk diameter however, the demagnetizing field driven deformation is significant. As already shown, the change in the net direction of this field when the core is displaced aligns with the core polarity, thus leading to an increase in the average out-of-plane magnetization of the displaced vortex state. Indeed for the case of $H/\hs=0.85$ we see that $\bar{m}_z$ becomes positive as the core oscillation amplitude increases with time. This is consistent with  core deformation  predominantly coming from demagnetizing field effects.

\section{Gyroconstant and stiffness coefficient calculations}
\label{scalcs}

In the previous section it was shown that for disks of small lateral dimensions in large out-of-plane fields, the demagnetizing field created when the vortex core is shifted laterally leads to significant deformation of the magnetization profile. Moreover, this high field deformation was shown to result in an increase in the average out-of-plane magnetization of the displaced vortex state. In this section we will relate this deformation to the non-linear field dependence of the gyrotropic frequency observed in Fig.~\ref{fdropoff}(a). To do this we  analyze the field dependence of the vortex stiffness coefficient, $\kappa$, and the gyroconstant, $G$ which together enable a calculation of $\fg$ according to the equation \cite{Guslienko2002}:
\begin{equation}
2\pi\fg=\frac{\kappa}{G}.
\label{ekG}
\end{equation}
The gyroconstant describes the lateral force acting on the core as a result of the motion of the non-uniform magnetization configuration \cite{Thiele1973,Huber1982}. $G$ can be calculated from the thickness averaged spin structure using the equation:
\begin{equation}
G=\frac{\ms L}{\gamma}\iint_{A}\mathbf{m}\cdot \left (\frac{d\mathbf{m}}{dx}\times\frac{d\mathbf{m}}{dy}\right )dx dy
\label{eG}
\end{equation} 
where for the small amplitude oscillations considered here the integration should be done over the area of the vortex core \cite{Fried2016}.

In Eq.~(\ref{ekG}), the stiffness coefficient describes the resorting force acting on the vortex core due to an increase in magnetic energy when it is displaced. For small shifts of the vortex core one can assume parabolic scaling of the magnetic energy with core displacement \cite{Guslienko2002,Dussaux2013,Sukhostavets2013}: 
\begin{equation}
W(X)=W(0)+\frac{1}{2}\kappa X^2+\mathcal{O}(X^4)
\label{ek}
\end{equation}
where $W$ is the total magnetic energy and $X=|\mathbf{X}|$ is the core displacement. Note that the Zeeman energy associated with the in-plane driving field should not be taken into account when determining $\kappa$. This is due to the fact that the in-plane field is only a mechanism to translate the core within the disk, thus enabling the confining potential of the vortex to be probed. 

\subsection{Field dependence of the gyroconstant}

The influence of a spatially uniform out-of-plane field on $G$ has previously been analytically studied\cite{deLoubens2009}. There, the gyroconstant was calculated using the expression:
\begin{equation}
G(H)=G(0)(1-p\cos\theta)
\label{eGdeLoubens}
\end{equation}
where $\theta$ is the polar angle (i.e. down from the $z$ axis) in which the magnetization outside the core is tilted \cite{PigeauThesis}. The above equation shows that $G$ decreases as the curling magnetization tilts to align with the external field. Such behavior can be understood from the fact that out-of-plane canting of the curling spins decreases the spatial gradient of the magnetization within the vortex core leading to a smaller value of $G$.

As previously shown\cite{Fried2016}, we find good agreement between the gyroconstant calculated directly from Eq.~(\ref{eG}) and values of $G$ predicted by Eq.~(\ref{eGdeLoubens}) when $\theta$ is taken from the minimum of the magnetostatic halo. For example, in the case of the 192$\,$nm disk we observe a maximum discrepancy of $\approx$ 2\% across the range of applied fields. For this reason, in the following discussion all values of $G$ have been calculated using the simpler Eq.~(\ref{eGdeLoubens}) with $\theta$ extracted from the minimum of the magnetostatic halo.

\begin{figure}[t]
	\centering
	\includegraphics[width=7.5cm]{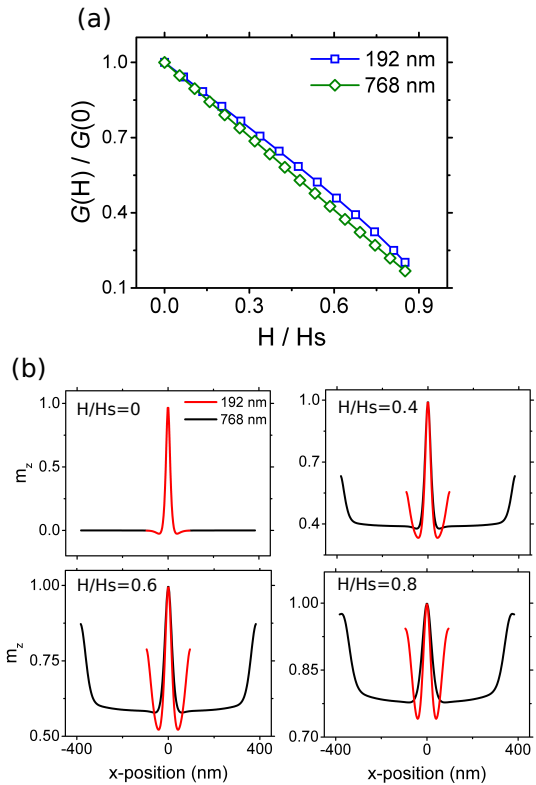}
	\caption{(a) Gyroconstant field dependence (normalised by its value in zero out-of-plane field). (b) Comparison of the static magnetization profiles of a centered vortex core for the 192 and 768 nm disks in various out-of-plane field amplitudes. The one dimensional slice has been taken across the disk center of the thickness averaged magnetization.}
	\label{fG}
\end{figure}

Resultant calculated values of the gyroconstant are shown for a range of out-of-plane field amplitudes in Fig.~\ref{fG}(a). Consistent with the above discussion, $G$ decreases with increasing field amplitude. One will note that  $G$ is consistently smaller for the larger disk diameter when in finite out-of-plane fields. To understand this behavior we compare the equilibrium out-of-plane magnetization configurations of the two disk sizes in a range of normalized field amplitudes [Fig.~\ref{fG}(b)]. Clearly there is a deepening of the magnetostatic halo for the smaller disk size when in a finite out-of-plane field. This is due to the nearby out-of-plane canting of the magnetization at the disk boundary which reinforces the negative demagnetizing field created by the vortex core (which is normally at the source of the halo \cite{Gaididei2010}). This halo deepening increases the spatial magnetization gradient within the vortex core resulting in a larger value of $G$ [as observed in Fig.~\ref{fG}(a)]. Nevertheless we note that for both disk sizes the gyroconstant continues to decrease with increasing field, a behavior which, ignoring field-dependencies of $\kappa$, would lead to a gyrotropic frequency which monotonically increases with out-of-plane field. Therefore to explain the drop-off in $\fg$ at high field amplitudes we must also to look at how $\kappa$ changes with increasing out-of-plane field. 

\subsection{Field dependence of the stiffness coefficient}

The influence of a uniform out-of-plane field on the stiffness coefficient was also studied in Ref.~\onlinecite{deLoubens2009}. This was done assuming that the only contribution to $\kappa$ was from the magnetostatic energy of the dipole charges that are generated by the in-plane magnetization when the core is shifted laterally. This leads to the expression for the field dependent stiffness coefficient:
\begin{equation}
\kappa(H)=\kappa(0)\sin^2\theta
\label{ekdeLoubens}
\end{equation}
showing that $\kappa$ also decreases as the magnetization cants to align with the applied field. Qualitatively this behavior can be understood from the fact that, as the out-of-plane field is increased there is a reduced in-plane magnetization component, meaning a weaker in-plane demagnetizing field is created by the curling spins when the vortex core is displaced. This results in a smaller increase in the system's magnetostatic energy when the core is displaced, leading to a lower value of $\kappa$.

If one assumes that the canting of the curling magnetization varies linearly with field amplitude (i.e. $\cos\theta=H/\hs$), Eqs.~(\ref{eGdeLoubens}) and ({\ref{ekdeLoubens}) combine to give \cite{deLoubens2009}:
\begin{equation}
\fg(H)=\fg(0)\left (1+p\frac{H}{\hs}\right ).
\label{efH}
\end{equation}
This suggests that the gyrotropic frequency will increase linearly with field amplitude up to the disk saturation field (at which point the gyrotropic mode can no longer be excited). 

To determine the stiffness coefficient from our simulations, the vortex core was driven by an in-plane sinusoidal field with amplitude 0.05$\,$mT and frequency equal to $\fg$ as found when analyzing the sinc pulse induced gyrotropic motion. This is necessary as the sinc pulse field used to determine $\fg$ induces very small amplitude oscillations in the core ($X<1\,$nm) which results in an increased numerical uncertainty in the extracted value of $\kappa$. The stiffness coefficient is then calculated by performing a parabolic fit to the total magnetic energy as a function of core displacement. It is again noted that the Zeeman energy associated with the in-plane driving field (which was calculated by multiplying the average in-plane magnetization by the driving field amplitude at that time), was not taken into account when performing such a fit. 

We also note that driving large amplitude oscillations of the vortex core can lead to non-linear behavior in the gyrotropic mode dynamics which can manifest as resonant peak splitting \cite{Buchanan2007} and peak foldover \cite{Buchanan2007,Drews2012,Guslienko2010}. However we have confirmed that the described sinusoidal field is not large enough to induce such behavior. This was done by plotting the maximum core displacement versus the driving frequency and comparing this to values of $\fg$ determined from the sinc excitation field. As observed in Fig.~\ref{fkappa}(a) the frequency resulting in the largest maximum core displacement closely corresponds to the resonant gyrotropic frequency found when the core is driven by a low amplitude transverse sinc field (as referenced by the vertical black dashed lines). 

\begin{figure}[t!]
	\centering
	\includegraphics[width=8.4cm]{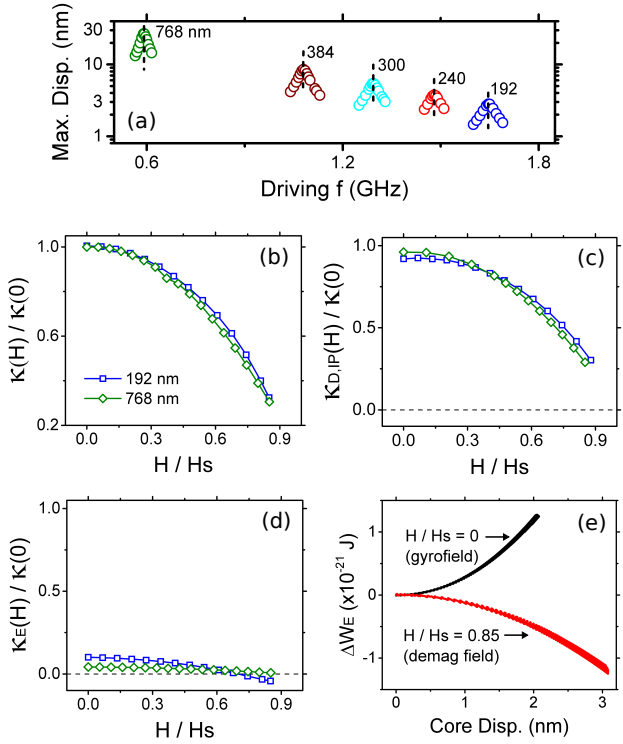}
	\caption{(a) Maximum core displacement as a function of sinusoidal driving frequency for a range of disk geometries (in a field close to $\hs$). (b) Total stiffness coefficient normalized by its value in zero out-of-plane field as a function of field amplitude for the 192 and 768$\,$nm disks. (c) The contribution to $\kappa$ arising from the magnetostatic energy associated with the in-plane magnetization and (d) the exchange energy. Both of these contributions have been normalized by the total stiffness coefficient in zero out-of-plane field. (e) The change in exchange energy (relative to that for a centered vortex) as a function of core displacement for the 192$\,$nm disk in $H/\hs$ = 0 and 0.85.}
	\label{fkappa}
\end{figure}

The resulting normalized values of $\kappa$ are shown for the two disk sizes in a range of field amplitudes in Fig.~\ref{fkappa}(b). We first note that, as expected from Eq.~(\ref{ekdeLoubens}), $\kappa$ decreases as the curling magnetization cants out-of-plane with increasing field amplitude. Such behavior in $\kappa$ acts to decrease $\fg$ (rather than increase it, as occurs in reality) showing that the linear frequency behavior observed for low out-of-plane fields is driven by changes in $G$. 

At intermediate field amplitudes $\kappa$ is larger for the smaller disk diameter. However, this increase in $\kappa$ is compensated  by a larger $G$ [as seen in Fig.~\ref{fG}(a)] resulting in similar normalized values of $\fg$ for the two disk sizes. As the field is further increased there is a sharper drop-off in $\kappa$ for the smaller disk size with the normalized stiffness coefficients of the two disk diameters becoming approximately equal close to $\hs$. This sharper drop-off in the stiffness coefficient of the smaller disk diameter is consistent with the non-linear frequency behavior observed in Fig.~\ref{fdropoff}(a). Namely this sharp drop-off leads to changes in $\kappa$ becoming comparable and then greater than changes in $G$, thus leading to a plateau and then drop-off in $\fg$ at high field amplitudes. In the next section we separate the individual contributions to the stiffness coefficient and demonstrate that this sharper drop-off in $\kappa$ is due to a decreased exchange contribution. 

\subsection{Stiffness coefficient decomposition}

The field behavior of the total vortex stiffness can be decomposed into two contributions. The first of these results from the magnetostatic  charges created by the in-plane magnetization when the vortex core is displaced and will be labeled $\kdip$. The second contribution, $\ke$, results from the change in exchange energy when the vortex core is displaced. $\ke$ and $\kdip$ were determined using an identical process to the total stiffness coefficient, however here only the appropriate energy contribution (as extracted from the simulations) is taken into account when performing the parabolic fit. It should be noted that we observed a non-zero contribution to $\kappa$ from the Zeeman energy associated with the out-of-plane magnetization, $\kz$, and the magnetostatic energy associated with the out-of-plane magnetization, $\kdoop$. However these two contributions cancel across the range of applied fields [$(\kdoop+\kz)/\kappa \leq 0.04$] and therefore have a negligible effect on $\fg$. 

Fig.~\ref{fkappa}(c) shows the field dependence of $\kdip$. Note that $\kdip$ has been normailzed by the value of the total stiffness coefficient in zero out-of-plane field. Consistent with the fact that Eq.~(\ref{ekdeLoubens}) was derived considering only the magnetostatic charges created by the in-plane magnetization when the vortex core is displaced, $\kdip$ varies as $\sin^2\theta$ to within 2\% for both disk sizes. We also note a sharper decrease in $\kdip$ for the 768$\,$nm disk as the field amplitude is increased. This is due to the fact that, as shown in Fig.~\ref{fG}(b), for finite field amplitudes the curling magnetization around the vortex core is less in-plane for the larger disk size (i.e.~the magnetostatic halo is higher). This means a weaker demagnetizing field is created by the in-plane magnetization when the vortex core is displaced, leading to a sharper decrease in $\kdip$. 

\begin{figure*}[t!]
	\centering
	\includegraphics[width=12cm]{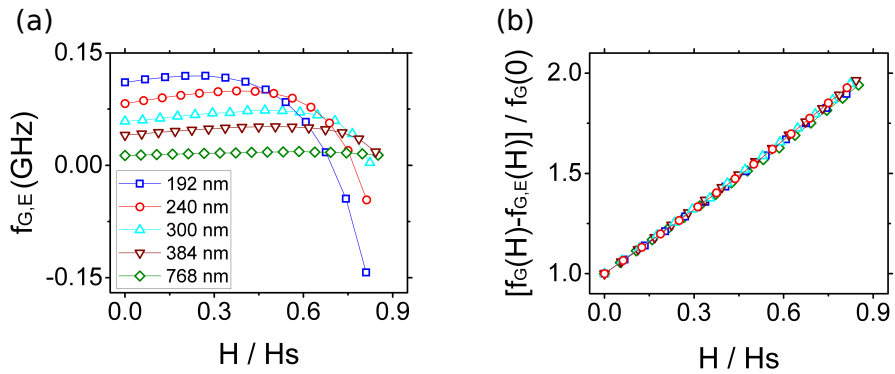}
	\caption{(a) The gyrotropic frequency associated with the exchange contribution to the stiffness coefficient for a range of disk diameters. (b) The simulated values of $\fg$ [as in Fig.~\ref{fdropoff}(a)] minus the frequency associated with the exchange contribution to the stiffness coefficient for a range of disk diameters.} 
	\label{fminusexch}
\end{figure*}

The field dependence of $\ke$ is shown in Fig.~\ref{fkappa}(d). For zero out-of-plane field, $\ke$ is positive for both disk sizes. $\ke$ then decreases as the out-of-plane field amplitude is increased. Notably this decrease is sharper for the smaller disk size, with $\ke$ even becoming negative for large $\hs$ (inferring that here the exchange energy favors core displacement). This sharper decrease $\ke$ is consistent with the more pronounced drop-off in the total stiffness coefficient for the smaller disk size when in large out-of-plane fields [Fig.~\ref{fkappa}(b)].

We have confirmed that the non-zero $\ke$ in zero field is primarily a result of vortex core deformation. This dominates the contribution to $\ke$ resulting from changes in the curling magnetization configuration when the core is displaced \cite{Guslienko2002}. This was demonstrated by calculating an exchange-driven core stiffness for statically displaced cores $\kappa_{\mathrm{E,static}}$. For the 192$\,$nm disk in zero out-of-plane field [where there is no deformation of the statically shifted core, i.e.~see Fig.~\ref{fdyncore}(a)] we found that $\kappa_{\mathrm{E,static}}$ was two orders of magnitude lower than the corresponding value of $\ke$ (the exchange-driven stiffness coefficient of the \emph{dynamic} core). This confirms that the dominant contribution to $\ke$ originates from vortex core deformation.

The out-of-plane field dependence of $\ke$ can therefore be understood by looking at how gyrofield-, and demagnetizing field-driven core deformation change the vortex exchange energy. This is done in Fig.~\ref{fkappa}(e) where we plot the change in exchange energy as a function of (dynamic) core displacement. We have done this for the 192$\,$nm disk in $H/\hs=0$  (i.e.~where deformation is driven by the gyrofield) and $H/\hs=0.85$ (where deformation is predominately a demagnetizing field effect). When in zero out-of-plane field, we see that the exchange energy increases as the core is displaced. This is consistent with the fact that the gyrofield decreases the average out-of-plane magnetization (Fig.~\ref{fdmz}). This results in the displaced vortex having a higher exchange energy than the centered vortex, since the magnetization is now further away from being uniformly saturated out-of-plane. In contrast, for the case of $H/\hs$ = 0.85 we see that the exchange energy decreases as the core is displaced. This is consistent with the fact that here deformation is driven by the demagnetizing field which results in an increase in the average out-of-plane magnetization (Fig.~\ref{fdmz}). This leads to the displaced vortex having a lower exchange energy than the centered vortex, since the magnetization is now closer to being uniformly saturated out-of-plane.

The field behavior of $\ke$ can now be qualitatively explained as follows. When in zero out-of-plane field $\ke$ is positive  for both disk sizes. This is consistent with the fact that here, the gyrofield (which generates a positive $\ke$) drives core deformation. For the larger disk size $\ke$ then decreases with increasing field amplitude due to a weakened gyrofield (Fig.~\ref{fgyrofield}). Indeed for large $H/\hs$, $\ke$ tends towards zero consistent with the fact that there is no strong core deformation [Fig.~\ref{fdyncore}(f)], due to a significantly reduced gyrofield. For the smaller disk size, $\ke$ also decreases with increasing field amplitude, however this decrease is sharper due to the presence of demagnetizing field driven core deformation (which generates a negative $\ke$). Indeed as previously noted, $\ke$ becomes negative at large out-of-plane fields, consistent with the fact that here core deformation is predominately a demagnetizing field effect. 

In Fig.~\ref{fminusexch} we show the field dependence of $\ke$ is the cause of the non-linear frequency behavior observed in Fig.~\ref{fdropoff}(a) for small disks in large out-of-plane fields. In Fig.~\ref{fminusexch}(a) we plot the field dependence of the frequency associated with the exchange contribution to the stiffness coefficient (i.e.~$f_{\mathrm{G,E}}=\ke/G$) for the range of disk geometries considered in Fig.~\ref{fdropoff}(a). For the larger disk size, $f_{\mathrm{G,E}}$ is almost independent of the field amplitude up to $\hs$. However for smaller disks, $f_{\mathrm{G,E}}$ is only constant at low field amplitudes, after which there is a sharp drop-off in the frequency. Consistent with the simulated frequency behavior in Fig.~\ref{fdropoff}(a), as the disk diameter is reduced, the decrease in $f_{\mathrm{G,E}}$ becomes sharper and begins at a lower field amplitude. On top of this, when subtracting $f_{\mathrm{G,E}}$ away from the simulated values of $\fg$, we find that for all disk sizes the resulting frequencies scale (almost) linearly up to the disk saturation field [Fig.~\ref{fminusexch}(b)]. This result explicitly confirms that the non-linear frequency behavior observed for small disks in Fig.~\ref{fdropoff}(a) is a result of a strongly decreasing $\ke$, which even becomes negative, in large out-of-plane fields. 

We finally note that it was found that the observed non-linear frequency behavior shows little dependence on the disk thickness. For example when comparing the normalized values of the gyrotropic frequency for disks of thickness $L$ = 15 and 30$\,$nm with a constant diameter of $2R$ = 192$\,$nm we see a maximum discrepancy of $\approx 2\%$ across the range of applied fields. This behavior is not surprising given that we have just shown that this non-linear behavior in $\fg$ for small disks in large out-of-plane fields is a result of a strongly decreasing $\ke$. This decrease in $\ke$ is a result of demagnetizing field driven vortex core deformation, which is present as a result of the close proximity of the vortex core to the disk's edge (and is therefore largely independent of the element thickness).

\section{Conclusion}
\label{sconc}

In conclusion we have shown that disks with small lateral dimensions exhibit an increasingly non-linear field dependence of the gyrotropic mode frequency in large out-of-plane spatially uniform magnetic fields. This non-linear frequency behavior was found to be associated with an increased deformation of the vortex core which arises for small disks in large out-of-plane fields due to asymmetries in the demagnetizing field profile when the core is shifted laterally. This form of deformation increases the average out-of-plane magnetization of the displaced vortex state and thus leads to a decrease in the exchange energy when the vortex core is shifted laterally. This results in a sharper decrease in the total vortex stiffness coefficient, which leads to the observed mode-softening or drop-off in $\fg$ for small magnetic disks in large out-of-plane fields. 

\begin{acknowledgments}

This work was supported by the Australian Research Council's Discovery Early Career Researcher Award scheme (DE120100155), a research grant from the United States Air Force (Asian Office of Aerospace Research and Development, AOARD), the University of Western Australia's Early Career Researcher Fellowship Support and Research Collaboration Award schemes and by resources provided by the Pawsey Supercomputing Centre with funding from the Australian Government and the Government of Western Australia.  The authors thank Maximilian Albert and Rebecca L.~Carey for assistance with the Finmag package. 

\end{acknowledgments}


\begin{thebibliography}{46}%
\makeatletter
\providecommand \@ifxundefined [1]{%
 \@ifx{#1\undefined}
}%
\providecommand \@ifnum [1]{%
 \ifnum #1\expandafter \@firstoftwo
 \else \expandafter \@secondoftwo
 \fi
}%
\providecommand \@ifx [1]{%
 \ifx #1\expandafter \@firstoftwo
 \else \expandafter \@secondoftwo
 \fi
}%
\providecommand \natexlab [1]{#1}%
\providecommand \enquote  [1]{``#1''}%
\providecommand \bibnamefont  [1]{#1}%
\providecommand \bibfnamefont [1]{#1}%
\providecommand \citenamefont [1]{#1}%
\providecommand \href@noop [0]{\@secondoftwo}%
\providecommand \href [0]{\begingroup \@sanitize@url \@href}%
\providecommand \@href[1]{\@@startlink{#1}\@@href}%
\providecommand \@@href[1]{\endgroup#1\@@endlink}%
\providecommand \@sanitize@url [0]{\catcode `\\12\catcode `\$12\catcode
  `\&12\catcode `\#12\catcode `\^12\catcode `\_12\catcode `\%12\relax}%
\providecommand \@@startlink[1]{}%
\providecommand \@@endlink[0]{}%
\providecommand \url  [0]{\begingroup\@sanitize@url \@url }%
\providecommand \@url [1]{\endgroup\@href {#1}{\urlprefix }}%
\providecommand \urlprefix  [0]{URL }%
\providecommand \Eprint [0]{\href }%
\providecommand \doibase [0]{http://dx.doi.org/}%
\providecommand \selectlanguage [0]{\@gobble}%
\providecommand \bibinfo  [0]{\@secondoftwo}%
\providecommand \bibfield  [0]{\@secondoftwo}%
\providecommand \translation [1]{[#1]}%
\providecommand \BibitemOpen [0]{}%
\providecommand \bibitemStop [0]{}%
\providecommand \bibitemNoStop [0]{.\EOS\space}%
\providecommand \EOS [0]{\spacefactor3000\relax}%
\providecommand \BibitemShut  [1]{\csname bibitem#1\endcsname}%
\let\auto@bib@innerbib\@empty
%</preamble>
\bibitem [{\citenamefont {Ha}\ \emph {et~al.}(2003)\citenamefont {Ha},
  \citenamefont {Hertel},\ and\ \citenamefont {Kirschner}}]{Ha2003}%
  \BibitemOpen
  \bibfield  {author} {\bibinfo {author} {\bibfnamefont {J.~K.}\ \bibnamefont
  {Ha}}, \bibinfo {author} {\bibfnamefont {R.}~\bibnamefont {Hertel}}, \ and\
  \bibinfo {author} {\bibfnamefont {J.}~\bibnamefont {Kirschner}},\ }\href
  {\doibase 10.1103/PhysRevB.67.224432} {\bibfield  {journal} {\bibinfo
  {journal} {Phys. Rev. B}\ }\textbf {\bibinfo {volume} {67}},\ \bibinfo
  {pages} {224432} (\bibinfo {year} {2003})}\BibitemShut {NoStop}%
\bibitem [{\citenamefont {Guslienko}\ and\ \citenamefont
  {Novosad}(2004)}]{Guslienko2004}%
  \BibitemOpen
  \bibfield  {author} {\bibinfo {author} {\bibfnamefont {K.~Y.}\ \bibnamefont
  {Guslienko}}\ and\ \bibinfo {author} {\bibfnamefont {V.}~\bibnamefont
  {Novosad}},\ }\href@noop {} {\bibfield  {journal} {\bibinfo  {journal} {J.
  Appl. Phys}\ }\textbf {\bibinfo {volume} {96}},\ \bibinfo {pages} {445}
  (\bibinfo {year} {2004})}\BibitemShut {NoStop}%
\bibitem [{\citenamefont {Cowburn}\ \emph {et~al.}(1999)\citenamefont
  {Cowburn}, \citenamefont {Koltsov}, \citenamefont {Adeyeye},\ and\
  \citenamefont {Welland}}]{Cowburn1999-2}%
  \BibitemOpen
  \bibfield  {author} {\bibinfo {author} {\bibfnamefont {R.~P.}\ \bibnamefont
  {Cowburn}}, \bibinfo {author} {\bibfnamefont {D.~K.}\ \bibnamefont
  {Koltsov}}, \bibinfo {author} {\bibfnamefont {A.~O.}\ \bibnamefont
  {Adeyeye}}, \ and\ \bibinfo {author} {\bibfnamefont {M.~E.}\ \bibnamefont
  {Welland}},\ }\href@noop {} {\bibfield  {journal} {\bibinfo  {journal} {Phys.
  Rev. Lett.}\ }\textbf {\bibinfo {volume} {83}},\ \bibinfo {pages} {1042}
  (\bibinfo {year} {1999})}\BibitemShut {NoStop}%
\bibitem [{\citenamefont {Shinjo}\ \emph {et~al.}(2000)\citenamefont {Shinjo},
  \citenamefont {Okuno}, \citenamefont {Hassdorf}, \citenamefont {Shigeto},\
  and\ \citenamefont {Ono}}]{Shinjo2000}%
  \BibitemOpen
  \bibfield  {author} {\bibinfo {author} {\bibfnamefont {T.}~\bibnamefont
  {Shinjo}}, \bibinfo {author} {\bibfnamefont {T.}~\bibnamefont {Okuno}},
  \bibinfo {author} {\bibfnamefont {R.}~\bibnamefont {Hassdorf}}, \bibinfo
  {author} {\bibfnamefont {K.}~\bibnamefont {Shigeto}}, \ and\ \bibinfo
  {author} {\bibfnamefont {T.}~\bibnamefont {Ono}},\ }\href@noop {} {\bibfield
  {journal} {\bibinfo  {journal} {Science}\ }\textbf {\bibinfo {volume}
  {289}},\ \bibinfo {pages} {930} (\bibinfo {year} {2000})}\BibitemShut
  {NoStop}%
\bibitem [{\citenamefont {Wachowiak}\ \emph {et~al.}(2002)\citenamefont
  {Wachowiak}, \citenamefont {Wiebe}, \citenamefont {Bode}, \citenamefont
  {Pietzsch}, \citenamefont {Morgenstern},\ and\ \citenamefont
  {Wiesendanger}}]{Wachowiak2002}%
  \BibitemOpen
  \bibfield  {author} {\bibinfo {author} {\bibfnamefont {A.}~\bibnamefont
  {Wachowiak}}, \bibinfo {author} {\bibfnamefont {J.}~\bibnamefont {Wiebe}},
  \bibinfo {author} {\bibfnamefont {M.}~\bibnamefont {Bode}}, \bibinfo {author}
  {\bibfnamefont {O.}~\bibnamefont {Pietzsch}}, \bibinfo {author}
  {\bibfnamefont {M.}~\bibnamefont {Morgenstern}}, \ and\ \bibinfo {author}
  {\bibfnamefont {R.}~\bibnamefont {Wiesendanger}},\ }\href {\doibase
  10.1126/science.1075302} {\bibfield  {journal} {\bibinfo  {journal}
  {Science}\ }\textbf {\bibinfo {volume} {298}},\ \bibinfo {pages} {577}
  (\bibinfo {year} {2002})}\BibitemShut {NoStop}%
\bibitem [{\citenamefont {Ding}\ \emph {et~al.}(2014)\citenamefont {Ding},
  \citenamefont {Kakazei}, \citenamefont {Liu}, \citenamefont {Guslienko},\
  and\ \citenamefont {Adeyeye}}]{Ding2014}%
  \BibitemOpen
  \bibfield  {author} {\bibinfo {author} {\bibfnamefont {J.}~\bibnamefont
  {Ding}}, \bibinfo {author} {\bibfnamefont {G.~N.}\ \bibnamefont {Kakazei}},
  \bibinfo {author} {\bibfnamefont {X.}~\bibnamefont {Liu}}, \bibinfo {author}
  {\bibfnamefont {K.~Y.}\ \bibnamefont {Guslienko}}, \ and\ \bibinfo {author}
  {\bibfnamefont {A.~O.}\ \bibnamefont {Adeyeye}},\ }\href@noop {} {\bibfield
  {journal} {\bibinfo  {journal} {Sci. Rep.}\ }\textbf {\bibinfo {volume}
  {4}},\ \bibinfo {pages} {4796} (\bibinfo {year} {2014})}\BibitemShut
  {NoStop}%
\bibitem [{\citenamefont {Kammerer}\ \emph {et~al.}(2011)\citenamefont
  {Kammerer}, \citenamefont {Weigand}, \citenamefont {Curcic}, \citenamefont
  {Noske}, \citenamefont {Sproll}, \citenamefont {Vansteenkiste}, \citenamefont
  {Van~Waeyenberge}, \citenamefont {Stoll}, \citenamefont {Woltersdorf},
  \citenamefont {Back},\ and\ \citenamefont {Schuetz}}]{Kammerer2011}%
  \BibitemOpen
  \bibfield  {author} {\bibinfo {author} {\bibfnamefont {M.}~\bibnamefont
  {Kammerer}}, \bibinfo {author} {\bibfnamefont {M.}~\bibnamefont {Weigand}},
  \bibinfo {author} {\bibfnamefont {M.}~\bibnamefont {Curcic}}, \bibinfo
  {author} {\bibfnamefont {M.}~\bibnamefont {Noske}}, \bibinfo {author}
  {\bibfnamefont {M.}~\bibnamefont {Sproll}}, \bibinfo {author} {\bibfnamefont
  {A.}~\bibnamefont {Vansteenkiste}}, \bibinfo {author} {\bibfnamefont
  {B.}~\bibnamefont {Van~Waeyenberge}}, \bibinfo {author} {\bibfnamefont
  {H.}~\bibnamefont {Stoll}}, \bibinfo {author} {\bibfnamefont
  {G.}~\bibnamefont {Woltersdorf}}, \bibinfo {author} {\bibfnamefont {C.~H.}\
  \bibnamefont {Back}}, \ and\ \bibinfo {author} {\bibfnamefont
  {G.}~\bibnamefont {Schuetz}},\ }\href {\doibase 10.1038/ncomms1277}
  {\bibfield  {journal} {\bibinfo  {journal} {Nat. Commun.}\ }\textbf {\bibinfo
  {volume} {2}},\ \bibinfo {pages} {279} (\bibinfo {year} {2011})}\BibitemShut
  {NoStop}%
\bibitem [{\citenamefont {Aliev}\ \emph {et~al.}(2009)\citenamefont {Aliev},
  \citenamefont {Sierra}, \citenamefont {Awad}, \citenamefont {Kakazei},
  \citenamefont {Han}, \citenamefont {Kim}, \citenamefont {Metlushko},
  \citenamefont {Ilic},\ and\ \citenamefont {Guslienko}}]{Aliev2009}%
  \BibitemOpen
  \bibfield  {author} {\bibinfo {author} {\bibfnamefont {F.~G.}\ \bibnamefont
  {Aliev}}, \bibinfo {author} {\bibfnamefont {J.~F.}\ \bibnamefont {Sierra}},
  \bibinfo {author} {\bibfnamefont {A.~A.}\ \bibnamefont {Awad}}, \bibinfo
  {author} {\bibfnamefont {G.~N.}\ \bibnamefont {Kakazei}}, \bibinfo {author}
  {\bibfnamefont {D.-S.}\ \bibnamefont {Han}}, \bibinfo {author} {\bibfnamefont
  {S.-K.}\ \bibnamefont {Kim}}, \bibinfo {author} {\bibfnamefont
  {V.}~\bibnamefont {Metlushko}}, \bibinfo {author} {\bibfnamefont
  {B.}~\bibnamefont {Ilic}}, \ and\ \bibinfo {author} {\bibfnamefont {K.~Y.}\
  \bibnamefont {Guslienko}},\ }\href@noop {} {\bibfield  {journal} {\bibinfo
  {journal} {Phys. Rev. B}\ }\textbf {\bibinfo {volume} {79}},\ \bibinfo
  {pages} {174433} (\bibinfo {year} {2009})}\BibitemShut {NoStop}%
\bibitem [{\citenamefont {Boust}\ and\ \citenamefont
  {Vukadinovic}(2004)}]{Boust2004}%
  \BibitemOpen
  \bibfield  {author} {\bibinfo {author} {\bibfnamefont {F.}~\bibnamefont
  {Boust}}\ and\ \bibinfo {author} {\bibfnamefont {N.}~\bibnamefont
  {Vukadinovic}},\ }\href@noop {} {\bibfield  {journal} {\bibinfo  {journal}
  {Phys. Rev. B}\ }\textbf {\bibinfo {volume} {70}},\ \bibinfo {pages} {172408}
  (\bibinfo {year} {2004})}\BibitemShut {NoStop}%
\bibitem [{\citenamefont {Ivanov}\ and\ \citenamefont
  {Zaspel}(2005)}]{Ivanov2005}%
  \BibitemOpen
  \bibfield  {author} {\bibinfo {author} {\bibfnamefont {B.~A.}\ \bibnamefont
  {Ivanov}}\ and\ \bibinfo {author} {\bibfnamefont {C.}~\bibnamefont
  {Zaspel}},\ }\href@noop {} {\bibfield  {journal} {\bibinfo  {journal} {Phys.
  Rev. Lett.}\ }\textbf {\bibinfo {volume} {94}},\ \bibinfo {pages} {027205}
  (\bibinfo {year} {2005})}\BibitemShut {NoStop}%
\bibitem [{\citenamefont {Choe}\ \emph {et~al.}(2004)\citenamefont {Choe},
  \citenamefont {Acremann}, \citenamefont {Scholl}, \citenamefont {Bauer},
  \citenamefont {Doran}, \citenamefont {St\"ohr},\ and\ \citenamefont
  {Padmore}}]{Choe2004}%
  \BibitemOpen
  \bibfield  {author} {\bibinfo {author} {\bibfnamefont {S.-B.}\ \bibnamefont
  {Choe}}, \bibinfo {author} {\bibfnamefont {Y.}~\bibnamefont {Acremann}},
  \bibinfo {author} {\bibfnamefont {A.}~\bibnamefont {Scholl}}, \bibinfo
  {author} {\bibfnamefont {A.}~\bibnamefont {Bauer}}, \bibinfo {author}
  {\bibfnamefont {A.}~\bibnamefont {Doran}}, \bibinfo {author} {\bibfnamefont
  {J.}~\bibnamefont {St\"ohr}}, \ and\ \bibinfo {author} {\bibfnamefont
  {H.~A.}\ \bibnamefont {Padmore}},\ }\href {\doibase 10.1126/science.1095068}
  {\bibfield  {journal} {\bibinfo  {journal} {Science}\ }\textbf {\bibinfo
  {volume} {304}},\ \bibinfo {pages} {420} (\bibinfo {year}
  {2004})}\BibitemShut {NoStop}%
\bibitem [{\citenamefont {Guslienko}\ \emph {et~al.}(2002)\citenamefont
  {Guslienko}, \citenamefont {Ivanov}, \citenamefont {Novosad}, \citenamefont
  {Otani}, \citenamefont {Shima},\ and\ \citenamefont
  {Fukamichi}}]{Guslienko2002}%
  \BibitemOpen
  \bibfield  {author} {\bibinfo {author} {\bibfnamefont {K.~Y.}\ \bibnamefont
  {Guslienko}}, \bibinfo {author} {\bibfnamefont {B.~A.}\ \bibnamefont
  {Ivanov}}, \bibinfo {author} {\bibfnamefont {V.}~\bibnamefont {Novosad}},
  \bibinfo {author} {\bibfnamefont {Y.}~\bibnamefont {Otani}}, \bibinfo
  {author} {\bibfnamefont {H.}~\bibnamefont {Shima}}, \ and\ \bibinfo {author}
  {\bibfnamefont {K.}~\bibnamefont {Fukamichi}},\ }\href {\doibase
  10.1063/1.1450816} {\bibfield  {journal} {\bibinfo  {journal} {J. Appl.
  Phys.}\ }\textbf {\bibinfo {volume} {91}},\ \bibinfo {pages} {8037} (\bibinfo
  {year} {2002})}\BibitemShut {NoStop}%
\bibitem [{\citenamefont {Park}\ \emph {et~al.}(2003)\citenamefont {Park},
  \citenamefont {Eames}, \citenamefont {Engebretson}, \citenamefont
  {Berezovsky},\ and\ \citenamefont {Crowell}}]{Park2003}%
  \BibitemOpen
  \bibfield  {author} {\bibinfo {author} {\bibfnamefont {J.~P.}\ \bibnamefont
  {Park}}, \bibinfo {author} {\bibfnamefont {P.}~\bibnamefont {Eames}},
  \bibinfo {author} {\bibfnamefont {D.~M.}\ \bibnamefont {Engebretson}},
  \bibinfo {author} {\bibfnamefont {J.}~\bibnamefont {Berezovsky}}, \ and\
  \bibinfo {author} {\bibfnamefont {P.~A.}\ \bibnamefont {Crowell}},\ }\href
  {\doibase 10.1103/PhysRevB.67.020403} {\bibfield  {journal} {\bibinfo
  {journal} {Phys. Rev. B}\ }\textbf {\bibinfo {volume} {67}},\ \bibinfo
  {pages} {020403} (\bibinfo {year} {2003})}\BibitemShut {NoStop}%
\bibitem [{\citenamefont {de~Loubens}\ \emph {et~al.}(2009)\citenamefont
  {de~Loubens}, \citenamefont {Riegler}, \citenamefont {Pigeau}, \citenamefont
  {Lochner}, \citenamefont {Boust}, \citenamefont {Guslienko}, \citenamefont
  {Hurdequint}, \citenamefont {Molenkamp}, \citenamefont {Schmidt},
  \citenamefont {Slavin}, \citenamefont {Tiberkevich}, \citenamefont
  {Vukadinovic},\ and\ \citenamefont {Klein}}]{deLoubens2009}%
  \BibitemOpen
  \bibfield  {author} {\bibinfo {author} {\bibfnamefont {G.}~\bibnamefont
  {de~Loubens}}, \bibinfo {author} {\bibfnamefont {A.}~\bibnamefont {Riegler}},
  \bibinfo {author} {\bibfnamefont {B.}~\bibnamefont {Pigeau}}, \bibinfo
  {author} {\bibfnamefont {F.}~\bibnamefont {Lochner}}, \bibinfo {author}
  {\bibfnamefont {F.}~\bibnamefont {Boust}}, \bibinfo {author} {\bibfnamefont
  {K.~Y.}\ \bibnamefont {Guslienko}}, \bibinfo {author} {\bibfnamefont
  {H.}~\bibnamefont {Hurdequint}}, \bibinfo {author} {\bibfnamefont {L.~W.}\
  \bibnamefont {Molenkamp}}, \bibinfo {author} {\bibfnamefont {G.}~\bibnamefont
  {Schmidt}}, \bibinfo {author} {\bibfnamefont {A.~N.}\ \bibnamefont {Slavin}},
  \bibinfo {author} {\bibfnamefont {V.~S.}\ \bibnamefont {Tiberkevich}},
  \bibinfo {author} {\bibfnamefont {N.}~\bibnamefont {Vukadinovic}}, \ and\
  \bibinfo {author} {\bibfnamefont {O.}~\bibnamefont {Klein}},\ }\href
  {\doibase 10.1103/PhysRevLett.102.177602} {\bibfield  {journal} {\bibinfo
  {journal} {Phys. Rev. Lett.}\ }\textbf {\bibinfo {volume} {102}},\ \bibinfo
  {pages} {177602} (\bibinfo {year} {2009})}\BibitemShut {NoStop}%
\bibitem [{\citenamefont {Yoo}\ \emph {et~al.}(2011)\citenamefont {Yoo},
  \citenamefont {Lee}, \citenamefont {Han},\ and\ \citenamefont
  {Kim}}]{Yoo2011}%
  \BibitemOpen
  \bibfield  {author} {\bibinfo {author} {\bibfnamefont {M.-W.}\ \bibnamefont
  {Yoo}}, \bibinfo {author} {\bibfnamefont {K.-S.}\ \bibnamefont {Lee}},
  \bibinfo {author} {\bibfnamefont {D.-S.}\ \bibnamefont {Han}}, \ and\
  \bibinfo {author} {\bibfnamefont {S.-K.}\ \bibnamefont {Kim}},\ }\href@noop
  {} {\bibfield  {journal} {\bibinfo  {journal} {J. Appl. Phys.}\ }\textbf
  {\bibinfo {volume} {109}},\ \bibinfo {pages} {063903} (\bibinfo {year}
  {2011})}\BibitemShut {NoStop}%
\bibitem [{\citenamefont {Buchanan}\ \emph {et~al.}(2006)\citenamefont
  {Buchanan}, \citenamefont {Roy}, \citenamefont {Grimsditch}, \citenamefont
  {Fradin}, \citenamefont {Guslienko}, \citenamefont {Bader},\ and\
  \citenamefont {Novosad}}]{Buchanan2006}%
  \BibitemOpen
  \bibfield  {author} {\bibinfo {author} {\bibfnamefont {K.~S.}\ \bibnamefont
  {Buchanan}}, \bibinfo {author} {\bibfnamefont {P.~E.}\ \bibnamefont {Roy}},
  \bibinfo {author} {\bibfnamefont {M.}~\bibnamefont {Grimsditch}}, \bibinfo
  {author} {\bibfnamefont {F.~Y.}\ \bibnamefont {Fradin}}, \bibinfo {author}
  {\bibfnamefont {K.~Y.}\ \bibnamefont {Guslienko}}, \bibinfo {author}
  {\bibfnamefont {S.~D.}\ \bibnamefont {Bader}}, \ and\ \bibinfo {author}
  {\bibfnamefont {V.}~\bibnamefont {Novosad}},\ }\href@noop {} {\bibfield
  {journal} {\bibinfo  {journal} {Phys. Rev. B}\ }\textbf {\bibinfo {volume}
  {74}},\ \bibinfo {pages} {064404} (\bibinfo {year} {2006})}\BibitemShut
  {NoStop}%
\bibitem [{\citenamefont {Yakata}\ \emph {et~al.}(2013)\citenamefont {Yakata},
  \citenamefont {Tanaka}, \citenamefont {Kiseki}, \citenamefont {Matsuyama},\
  and\ \citenamefont {Kimura}}]{Yakata2013}%
  \BibitemOpen
  \bibfield  {author} {\bibinfo {author} {\bibfnamefont {S.}~\bibnamefont
  {Yakata}}, \bibinfo {author} {\bibfnamefont {T.}~\bibnamefont {Tanaka}},
  \bibinfo {author} {\bibfnamefont {K.}~\bibnamefont {Kiseki}}, \bibinfo
  {author} {\bibfnamefont {K.}~\bibnamefont {Matsuyama}}, \ and\ \bibinfo
  {author} {\bibfnamefont {T.}~\bibnamefont {Kimura}},\ }\href
  {http://dx.doi.org/10.1038/srep03567} {\bibfield  {journal} {\bibinfo
  {journal} {Sci. Rep.}\ }\textbf {\bibinfo {volume} {3}},\ \bibinfo {pages}
  {3567} (\bibinfo {year} {2013})}\BibitemShut {NoStop}%
\bibitem [{\citenamefont {Ivanov}\ and\ \citenamefont
  {Zaspel}(2002)}]{Ivanov2002}%
  \BibitemOpen
  \bibfield  {author} {\bibinfo {author} {\bibfnamefont {B.~A.}\ \bibnamefont
  {Ivanov}}\ and\ \bibinfo {author} {\bibfnamefont {C.~E.}\ \bibnamefont
  {Zaspel}},\ }\href {\doibase 10.1063/1.1499515} {\bibfield  {journal}
  {\bibinfo  {journal} {Appl. Phys. Lett.}\ }\textbf {\bibinfo {volume} {81}},\
  \bibinfo {pages} {1261} (\bibinfo {year} {2002})}\BibitemShut {NoStop}%
\bibitem [{\citenamefont {Dussaux}\ \emph {et~al.}(2010)\citenamefont
  {Dussaux}, \citenamefont {Georges}, \citenamefont {Grollier}, \citenamefont
  {Cros}, \citenamefont {Khvalkovskiy}, \citenamefont {Fukushima},
  \citenamefont {Konoto}, \citenamefont {Kubota}, \citenamefont {Yakushiji},
  \citenamefont {Yuasa}, \citenamefont {Zvezdin}, \citenamefont {Ando},\ and\
  \citenamefont {Fert}}]{Dussaux2010}%
  \BibitemOpen
  \bibfield  {author} {\bibinfo {author} {\bibfnamefont {A.}~\bibnamefont
  {Dussaux}}, \bibinfo {author} {\bibfnamefont {B.}~\bibnamefont {Georges}},
  \bibinfo {author} {\bibfnamefont {J.}~\bibnamefont {Grollier}}, \bibinfo
  {author} {\bibfnamefont {V.}~\bibnamefont {Cros}}, \bibinfo {author}
  {\bibfnamefont {A.~V.}\ \bibnamefont {Khvalkovskiy}}, \bibinfo {author}
  {\bibfnamefont {A.}~\bibnamefont {Fukushima}}, \bibinfo {author}
  {\bibfnamefont {M.}~\bibnamefont {Konoto}}, \bibinfo {author} {\bibfnamefont
  {H.}~\bibnamefont {Kubota}}, \bibinfo {author} {\bibfnamefont
  {K.}~\bibnamefont {Yakushiji}}, \bibinfo {author} {\bibfnamefont
  {S.}~\bibnamefont {Yuasa}}, \bibinfo {author} {\bibfnamefont {K.~A.}\
  \bibnamefont {Zvezdin}}, \bibinfo {author} {\bibfnamefont {K.}~\bibnamefont
  {Ando}}, \ and\ \bibinfo {author} {\bibfnamefont {A.}~\bibnamefont {Fert}},\
  }\href@noop {} {\bibfield  {journal} {\bibinfo  {journal} {Nat. Commun.}\
  }\textbf {\bibinfo {volume} {1}},\ \bibinfo {pages} {8} (\bibinfo {year}
  {2010})}\BibitemShut {NoStop}%
\bibitem [{\citenamefont {Lebrun}\ \emph {et~al.}(2014)\citenamefont {Lebrun},
  \citenamefont {Locatelli}, \citenamefont {Tsunegi}, \citenamefont {Grollier},
  \citenamefont {Cros}, \citenamefont {Abreu~Araujo}, \citenamefont {Kubota},
  \citenamefont {Yakushiji}, \citenamefont {Fukushima},\ and\ \citenamefont
  {Yuasa}}]{Lebrun2014}%
  \BibitemOpen
  \bibfield  {author} {\bibinfo {author} {\bibfnamefont {R.}~\bibnamefont
  {Lebrun}}, \bibinfo {author} {\bibfnamefont {N.}~\bibnamefont {Locatelli}},
  \bibinfo {author} {\bibfnamefont {S.}~\bibnamefont {Tsunegi}}, \bibinfo
  {author} {\bibfnamefont {J.}~\bibnamefont {Grollier}}, \bibinfo {author}
  {\bibfnamefont {V.}~\bibnamefont {Cros}}, \bibinfo {author} {\bibfnamefont
  {F.}~\bibnamefont {Abreu~Araujo}}, \bibinfo {author} {\bibfnamefont
  {H.}~\bibnamefont {Kubota}}, \bibinfo {author} {\bibfnamefont
  {K.}~\bibnamefont {Yakushiji}}, \bibinfo {author} {\bibfnamefont
  {A.}~\bibnamefont {Fukushima}}, \ and\ \bibinfo {author} {\bibfnamefont
  {S.}~\bibnamefont {Yuasa}},\ }\href@noop {} {\bibfield  {journal} {\bibinfo
  {journal} {Phys. Rev. App.}\ }\textbf {\bibinfo {volume} {2}},\ \bibinfo
  {pages} {061001} (\bibinfo {year} {2014})}\BibitemShut {NoStop}%
\bibitem [{\citenamefont {Pribiag}\ \emph {et~al.}(2007)\citenamefont
  {Pribiag}, \citenamefont {Krivorotov}, \citenamefont {Fuchs}, \citenamefont
  {Braganca}, \citenamefont {Ozatay}, \citenamefont {Sankey}, \citenamefont
  {Ralph},\ and\ \citenamefont {Buhrman}}]{Pribiag2007}%
  \BibitemOpen
  \bibfield  {author} {\bibinfo {author} {\bibfnamefont {V.~S.}\ \bibnamefont
  {Pribiag}}, \bibinfo {author} {\bibfnamefont {I.~N.}\ \bibnamefont
  {Krivorotov}}, \bibinfo {author} {\bibfnamefont {G.~D.}\ \bibnamefont
  {Fuchs}}, \bibinfo {author} {\bibfnamefont {P.~M.}\ \bibnamefont {Braganca}},
  \bibinfo {author} {\bibfnamefont {O.}~\bibnamefont {Ozatay}}, \bibinfo
  {author} {\bibfnamefont {J.~C.}\ \bibnamefont {Sankey}}, \bibinfo {author}
  {\bibfnamefont {D.~C.}\ \bibnamefont {Ralph}}, \ and\ \bibinfo {author}
  {\bibfnamefont {R.~A.}\ \bibnamefont {Buhrman}},\ }\href@noop {} {\bibfield
  {journal} {\bibinfo  {journal} {Nat. Phys.}\ }\textbf {\bibinfo {volume}
  {3}},\ \bibinfo {pages} {498} (\bibinfo {year} {2007})}\BibitemShut {NoStop}%
\bibitem [{\citenamefont {Braganca}\ \emph {et~al.}(2010)\citenamefont
  {Braganca}, \citenamefont {Gurney}, \citenamefont {Wilson}, \citenamefont
  {Katine}, \citenamefont {Maat},\ and\ \citenamefont
  {Childress}}]{Braganca2010}%
  \BibitemOpen
  \bibfield  {author} {\bibinfo {author} {\bibfnamefont {P.~M.}\ \bibnamefont
  {Braganca}}, \bibinfo {author} {\bibfnamefont {B.~A.}\ \bibnamefont
  {Gurney}}, \bibinfo {author} {\bibfnamefont {B.~A.}\ \bibnamefont {Wilson}},
  \bibinfo {author} {\bibfnamefont {J.~A.}\ \bibnamefont {Katine}}, \bibinfo
  {author} {\bibfnamefont {S.}~\bibnamefont {Maat}}, \ and\ \bibinfo {author}
  {\bibfnamefont {J.~R.}\ \bibnamefont {Childress}},\ }\href@noop {} {\bibfield
   {journal} {\bibinfo  {journal} {Nanotechnol.}\ }\textbf {\bibinfo {volume}
  {21}},\ \bibinfo {pages} {235202} (\bibinfo {year} {2010})}\BibitemShut
  {NoStop}%
\bibitem [{\citenamefont {Fried}\ and\ \citenamefont
  {Metaxas}(2016)}]{Fried2016}%
  \BibitemOpen
  \bibfield  {author} {\bibinfo {author} {\bibfnamefont {J.~P.}\ \bibnamefont
  {Fried}}\ and\ \bibinfo {author} {\bibfnamefont {P.~J.}\ \bibnamefont
  {Metaxas}},\ }\href {\doibase 10.1103/PhysRevB.93.064422} {\bibfield
  {journal} {\bibinfo  {journal} {Phys. Rev. B}\ }\textbf {\bibinfo {volume}
  {93}},\ \bibinfo {pages} {064422} (\bibinfo {year} {2016})}\BibitemShut
  {NoStop}%
\bibitem [{\citenamefont {Dussaux}(2011)}]{DussauxThesis}%
  \BibitemOpen
  \bibfield  {author} {\bibinfo {author} {\bibfnamefont {A.}~\bibnamefont
  {Dussaux}},\ }\emph {\bibinfo {title} {Etude des oscillations de vortex
  magn{\'e}tiques induites par transfert de spin}},\ \href@noop {} {Ph.D.
  thesis},\ \bibinfo  {school} {Universite Paris VI} (\bibinfo {year}
  {2011})\BibitemShut {NoStop}%
\bibitem [{\citenamefont {Vansteenkiste}\ \emph {et~al.}(2014)\citenamefont
  {Vansteenkiste}, \citenamefont {Leliaert}, \citenamefont {Dvornik},
  \citenamefont {Helsen}, \citenamefont {Garcia-Sanchez},\ and\ \citenamefont
  {{Van Waeyenberg}}}]{Vansteenkiste2014}%
  \BibitemOpen
  \bibfield  {author} {\bibinfo {author} {\bibfnamefont {A.}~\bibnamefont
  {Vansteenkiste}}, \bibinfo {author} {\bibfnamefont {J.}~\bibnamefont
  {Leliaert}}, \bibinfo {author} {\bibfnamefont {M.}~\bibnamefont {Dvornik}},
  \bibinfo {author} {\bibfnamefont {M.}~\bibnamefont {Helsen}}, \bibinfo
  {author} {\bibfnamefont {F.}~\bibnamefont {Garcia-Sanchez}}, \ and\ \bibinfo
  {author} {\bibfnamefont {B.}~\bibnamefont {{Van Waeyenberg}}},\ }\href@noop
  {} {\bibfield  {journal} {\bibinfo  {journal} {AIP Adv.}\ }\textbf {\bibinfo
  {volume} {4}},\ \bibinfo {pages} {107133} (\bibinfo {year}
  {2014})}\BibitemShut {NoStop}%
\bibitem [{\citenamefont {Baker}\ \emph {et~al.}(2016)\citenamefont {Baker},
  \citenamefont {Beg}, \citenamefont {Asthon}, \citenamefont {Albert},
  \citenamefont {Chernyshenko}, \citenamefont {Wang}, \citenamefont {Zhang},
  \citenamefont {Bisotti}, \citenamefont {Franchin}, \citenamefont {Hu},
  \citenamefont {Stamps}, \citenamefont {Hesjedal},\ and\ \citenamefont
  {Fanhour}}]{Baker2016}%
  \BibitemOpen
  \bibfield  {author} {\bibinfo {author} {\bibfnamefont {A.}~\bibnamefont
  {Baker}}, \bibinfo {author} {\bibfnamefont {M.}~\bibnamefont {Beg}}, \bibinfo
  {author} {\bibfnamefont {G.}~\bibnamefont {Asthon}}, \bibinfo {author}
  {\bibfnamefont {G.}~\bibnamefont {Albert}}, \bibinfo {author} {\bibfnamefont
  {D.}~\bibnamefont {Chernyshenko}}, \bibinfo {author} {\bibfnamefont
  {W.}~\bibnamefont {Wang}}, \bibinfo {author} {\bibfnamefont {S.}~\bibnamefont
  {Zhang}}, \bibinfo {author} {\bibfnamefont {M.}~\bibnamefont {Bisotti}},
  \bibinfo {author} {\bibfnamefont {M.}~\bibnamefont {Franchin}}, \bibinfo
  {author} {\bibfnamefont {C.}~\bibnamefont {Hu}}, \bibinfo {author}
  {\bibfnamefont {R.}~\bibnamefont {Stamps}}, \bibinfo {author} {\bibfnamefont
  {T.}~\bibnamefont {Hesjedal}}, \ and\ \bibinfo {author} {\bibfnamefont
  {H.}~\bibnamefont {Fanhour}},\ }\href@noop {} {\bibfield  {journal} {\bibinfo
   {journal} {arXiv:1603.05419}\ } (\bibinfo {year} {2016})}\BibitemShut
  {NoStop}%
\bibitem [{\citenamefont {Park}\ and\ \citenamefont
  {Crowell}(2005)}]{Park2005}%
  \BibitemOpen
  \bibfield  {author} {\bibinfo {author} {\bibfnamefont {J.~P.}\ \bibnamefont
  {Park}}\ and\ \bibinfo {author} {\bibfnamefont {P.~A.}\ \bibnamefont
  {Crowell}},\ }\href@noop {} {\bibfield  {journal} {\bibinfo  {journal} {Phys.
  Rev. Lett.}\ }\textbf {\bibinfo {volume} {95}},\ \bibinfo {pages} {167201}
  (\bibinfo {year} {2005})}\BibitemShut {NoStop}%
\bibitem [{\citenamefont {Buess}\ \emph {et~al.}(2004)\citenamefont {Buess},
  \citenamefont {H{\"o}llinger}, \citenamefont {Haug}, \citenamefont
  {Perzlmaier}, \citenamefont {Krey}, \citenamefont {Pescia}, \citenamefont
  {Scheinfein}, \citenamefont {Weiss},\ and\ \citenamefont {Back}}]{Buess2004}%
  \BibitemOpen
  \bibfield  {author} {\bibinfo {author} {\bibfnamefont {M.}~\bibnamefont
  {Buess}}, \bibinfo {author} {\bibfnamefont {R.}~\bibnamefont
  {H{\"o}llinger}}, \bibinfo {author} {\bibfnamefont {T.}~\bibnamefont {Haug}},
  \bibinfo {author} {\bibfnamefont {K.}~\bibnamefont {Perzlmaier}}, \bibinfo
  {author} {\bibfnamefont {U.}~\bibnamefont {Krey}}, \bibinfo {author}
  {\bibfnamefont {D.}~\bibnamefont {Pescia}}, \bibinfo {author} {\bibfnamefont
  {M.~R.}\ \bibnamefont {Scheinfein}}, \bibinfo {author} {\bibfnamefont
  {D.}~\bibnamefont {Weiss}}, \ and\ \bibinfo {author} {\bibfnamefont {C.~H.}\
  \bibnamefont {Back}},\ }\href@noop {} {\bibfield  {journal} {\bibinfo
  {journal} {Phys. Rev. Lett.}\ }\textbf {\bibinfo {volume} {93}},\ \bibinfo
  {pages} {077207} (\bibinfo {year} {2004})}\BibitemShut {NoStop}%
\bibitem [{\citenamefont {Dussaux}\ \emph {et~al.}(2011)\citenamefont
  {Dussaux}, \citenamefont {Khvalkovskiy}, \citenamefont {Grollier},
  \citenamefont {Cros}, \citenamefont {Fukushima}, \citenamefont {Konoto},
  \citenamefont {Kubota}, \citenamefont {Yakushiji}, \citenamefont {Yuasa},
  \citenamefont {Ando},\ and\ \citenamefont {Fert}}]{Dussaux2011}%
  \BibitemOpen
  \bibfield  {author} {\bibinfo {author} {\bibfnamefont {A.}~\bibnamefont
  {Dussaux}}, \bibinfo {author} {\bibfnamefont {A.~V.}\ \bibnamefont
  {Khvalkovskiy}}, \bibinfo {author} {\bibfnamefont {J.}~\bibnamefont
  {Grollier}}, \bibinfo {author} {\bibfnamefont {V.}~\bibnamefont {Cros}},
  \bibinfo {author} {\bibfnamefont {A.}~\bibnamefont {Fukushima}}, \bibinfo
  {author} {\bibfnamefont {M.}~\bibnamefont {Konoto}}, \bibinfo {author}
  {\bibfnamefont {H.}~\bibnamefont {Kubota}}, \bibinfo {author} {\bibfnamefont
  {K.}~\bibnamefont {Yakushiji}}, \bibinfo {author} {\bibfnamefont
  {S.}~\bibnamefont {Yuasa}}, \bibinfo {author} {\bibfnamefont
  {K.}~\bibnamefont {Ando}}, \ and\ \bibinfo {author} {\bibfnamefont
  {A.}~\bibnamefont {Fert}},\ }\href {\doibase 10.1063/1.3565159} {\bibfield
  {journal} {\bibinfo  {journal} {Appl. Phys. Lett.}\ }\textbf {\bibinfo
  {volume} {98}},\ \bibinfo {eid} {132506} (\bibinfo {year}
  {2011})}\BibitemShut {NoStop}%
\bibitem [{\citenamefont {Donahue}\ and\ \citenamefont {Porter}(1999)}]{oommf}%
  \BibitemOpen
  \bibfield  {author} {\bibinfo {author} {\bibfnamefont {M.~J.}\ \bibnamefont
  {Donahue}}\ and\ \bibinfo {author} {\bibfnamefont {D.~G.}\ \bibnamefont
  {Porter}},\ }\href@noop {} {\emph {\bibinfo {title} {{OOMMF User's Guide,
  Version 1.0, Interagency Report NISTIR 6376}}}},\ \bibinfo {type} {Tech.
  Rep.}\ (\bibinfo  {institution} {National Institute of Standards and
  Technology},\ \bibinfo {address} {Gaithersburg, MD},\ \bibinfo {year}
  {1999})\BibitemShut {NoStop}%
\bibitem [{\citenamefont {{d'Aquino}}\ \emph {et~al.}(2009)\citenamefont
  {{d'Aquino}}, \citenamefont {Serpico}, \citenamefont {Miano},\ and\
  \citenamefont {Forestiere}}]{dAquino2009}%
  \BibitemOpen
  \bibfield  {author} {\bibinfo {author} {\bibfnamefont {M.}~\bibnamefont
  {{d'Aquino}}}, \bibinfo {author} {\bibfnamefont {C.}~\bibnamefont {Serpico}},
  \bibinfo {author} {\bibfnamefont {G.}~\bibnamefont {Miano}}, \ and\ \bibinfo
  {author} {\bibfnamefont {C.}~\bibnamefont {Forestiere}},\ }\href@noop {}
  {\bibfield  {journal} {\bibinfo  {journal} {J. Comp. Phys.}\ }\textbf
  {\bibinfo {volume} {228}},\ \bibinfo {pages} {6130} (\bibinfo {year}
  {2009})}\BibitemShut {NoStop}%
\bibitem [{\citenamefont {Metaxas}\ \emph {et~al.}(2016)\citenamefont
  {Metaxas}, \citenamefont {Albert}, \citenamefont {Lequeux}, \citenamefont
  {Cros}, \citenamefont {Grollier}, \citenamefont {Bortolotti}, \citenamefont
  {Anane},\ and\ \citenamefont {Fangohr}}]{Metaxas2014}%
  \BibitemOpen
  \bibfield  {author} {\bibinfo {author} {\bibfnamefont {P.~J.}\ \bibnamefont
  {Metaxas}}, \bibinfo {author} {\bibfnamefont {M.}~\bibnamefont {Albert}},
  \bibinfo {author} {\bibfnamefont {S.}~\bibnamefont {Lequeux}}, \bibinfo
  {author} {\bibfnamefont {V.}~\bibnamefont {Cros}}, \bibinfo {author}
  {\bibfnamefont {J.}~\bibnamefont {Grollier}}, \bibinfo {author}
  {\bibfnamefont {P.}~\bibnamefont {Bortolotti}}, \bibinfo {author}
  {\bibfnamefont {A.}~\bibnamefont {Anane}}, \ and\ \bibinfo {author}
  {\bibfnamefont {H.}~\bibnamefont {Fangohr}},\ }\href {\doibase
  10.1103/PhysRevB.93.054414} {\bibfield  {journal} {\bibinfo  {journal} {Phys.
  Rev. B}\ }\textbf {\bibinfo {volume} {93}},\ \bibinfo {pages} {054414}
  (\bibinfo {year} {2016})}\BibitemShut {NoStop}%
\bibitem [{\citenamefont {Fischbacher}\ \emph {et~al.}(2007)\citenamefont
  {Fischbacher}, \citenamefont {Franchin}, \citenamefont {Bordignon},\ and\
  \citenamefont {Fangohr}}]{Fischbacher2007}%
  \BibitemOpen
  \bibfield  {author} {\bibinfo {author} {\bibfnamefont {T.}~\bibnamefont
  {Fischbacher}}, \bibinfo {author} {\bibfnamefont {M.}~\bibnamefont
  {Franchin}}, \bibinfo {author} {\bibfnamefont {G.}~\bibnamefont {Bordignon}},
  \ and\ \bibinfo {author} {\bibfnamefont {H.}~\bibnamefont {Fangohr}},\
  }\href@noop {} {\bibfield  {journal} {\bibinfo  {journal} {IEEE Trans. Mag.}\
  }\textbf {\bibinfo {volume} {43}},\ \bibinfo {pages} {2896} (\bibinfo {year}
  {2007})}\BibitemShut {NoStop}%
\bibitem [{\citenamefont {Gaididei}\ \emph {et~al.}(2010)\citenamefont
  {Gaididei}, \citenamefont {Kravchuk},\ and\ \citenamefont
  {Sehka}}]{Gaididei2010}%
  \BibitemOpen
  \bibfield  {author} {\bibinfo {author} {\bibfnamefont {Y.}~\bibnamefont
  {Gaididei}}, \bibinfo {author} {\bibfnamefont {V.~P.}\ \bibnamefont
  {Kravchuk}}, \ and\ \bibinfo {author} {\bibfnamefont {D.~D.}\ \bibnamefont
  {Sehka}},\ }\href@noop {} {\bibfield  {journal} {\bibinfo  {journal} {Int. J.
  Quant. Chem.}\ }\textbf {\bibinfo {volume} {110}},\ \bibinfo {pages} {83}
  (\bibinfo {year} {2010})}\BibitemShut {NoStop}%
\bibitem [{\citenamefont {Yamada}\ \emph {et~al.}(2007)\citenamefont {Yamada},
  \citenamefont {Kasai}, \citenamefont {Nakatani}, \citenamefont {Kobayashi},
  \citenamefont {Kohno}, \citenamefont {Thiaville},\ and\ \citenamefont
  {Ono}}]{Yamada2007}%
  \BibitemOpen
  \bibfield  {author} {\bibinfo {author} {\bibfnamefont {K.}~\bibnamefont
  {Yamada}}, \bibinfo {author} {\bibfnamefont {S.}~\bibnamefont {Kasai}},
  \bibinfo {author} {\bibfnamefont {Y.}~\bibnamefont {Nakatani}}, \bibinfo
  {author} {\bibfnamefont {K.}~\bibnamefont {Kobayashi}}, \bibinfo {author}
  {\bibfnamefont {H.}~\bibnamefont {Kohno}}, \bibinfo {author} {\bibfnamefont
  {A.}~\bibnamefont {Thiaville}}, \ and\ \bibinfo {author} {\bibfnamefont
  {T.}~\bibnamefont {Ono}},\ }\href {\doibase 10.1038/nmat1867} {\bibfield
  {journal} {\bibinfo  {journal} {Nat. Mater.}\ }\textbf {\bibinfo {volume}
  {6}},\ \bibinfo {pages} {270} (\bibinfo {year} {2007})}\BibitemShut {NoStop}%
\bibitem [{\citenamefont {Van~Waeyenberge}\ \emph {et~al.}(2006)\citenamefont
  {Van~Waeyenberge}, \citenamefont {Puzic}, \citenamefont {Stoll},
  \citenamefont {Chou}, \citenamefont {Hertel}, \citenamefont {Fahnle},
  \citenamefont {Bruckl}, \citenamefont {Rott}, \citenamefont {Reiss},
  \citenamefont {Neudecker}, \citenamefont {Weiss}, \citenamefont {Back},\ and\
  \citenamefont {Schutz}}]{Waeyenberge2006}%
  \BibitemOpen
  \bibfield  {author} {\bibinfo {author} {\bibfnamefont {B.}~\bibnamefont
  {Van~Waeyenberge}}, \bibinfo {author} {\bibfnamefont {A.}~\bibnamefont
  {Puzic}}, \bibinfo {author} {\bibfnamefont {H.}~\bibnamefont {Stoll}},
  \bibinfo {author} {\bibfnamefont {T.}~\bibnamefont {Chou}, \bibfnamefont
  {K.and~Tyliszczak}}, \bibinfo {author} {\bibfnamefont {R.}~\bibnamefont
  {Hertel}}, \bibinfo {author} {\bibfnamefont {M.}~\bibnamefont {Fahnle}},
  \bibinfo {author} {\bibfnamefont {H.}~\bibnamefont {Bruckl}}, \bibinfo
  {author} {\bibfnamefont {K.}~\bibnamefont {Rott}}, \bibinfo {author}
  {\bibfnamefont {G.}~\bibnamefont {Reiss}}, \bibinfo {author} {\bibfnamefont
  {I.}~\bibnamefont {Neudecker}}, \bibinfo {author} {\bibfnamefont
  {D.}~\bibnamefont {Weiss}}, \bibinfo {author} {\bibfnamefont {C.~H.}\
  \bibnamefont {Back}}, \ and\ \bibinfo {author} {\bibfnamefont
  {G.}~\bibnamefont {Schutz}},\ }\href@noop {} {\bibfield  {journal} {\bibinfo
  {journal} {Nature}\ }\textbf {\bibinfo {volume} {444}},\ \bibinfo {pages}
  {461} (\bibinfo {year} {2006})}\BibitemShut {NoStop}%
\bibitem [{\citenamefont {Novosad}\ \emph {et~al.}(2005)\citenamefont
  {Novosad}, \citenamefont {Fradin}, \citenamefont {Roy}, \citenamefont
  {Buchanan}, \citenamefont {Guslienko},\ and\ \citenamefont
  {Bader}}]{Novosad2005}%
  \BibitemOpen
  \bibfield  {author} {\bibinfo {author} {\bibfnamefont {V.}~\bibnamefont
  {Novosad}}, \bibinfo {author} {\bibfnamefont {F.~Y.}\ \bibnamefont {Fradin}},
  \bibinfo {author} {\bibfnamefont {P.~E.}\ \bibnamefont {Roy}}, \bibinfo
  {author} {\bibfnamefont {K.~S.}\ \bibnamefont {Buchanan}}, \bibinfo {author}
  {\bibfnamefont {K.~Y.}\ \bibnamefont {Guslienko}}, \ and\ \bibinfo {author}
  {\bibfnamefont {S.~D.}\ \bibnamefont {Bader}},\ }\href@noop {} {\bibfield
  {journal} {\bibinfo  {journal} {Phys. Rev. B}\ }\textbf {\bibinfo {volume}
  {72}},\ \bibinfo {pages} {024455} (\bibinfo {year} {2005})},\ \bibinfo {note}
  {decra}\BibitemShut {NoStop}%
\bibitem [{\citenamefont {Guslienko}\ \emph {et~al.}(2008)\citenamefont
  {Guslienko}, \citenamefont {Lee},\ and\ \citenamefont
  {Kim}}]{Guslienko2008b}%
  \BibitemOpen
  \bibfield  {author} {\bibinfo {author} {\bibfnamefont {K.~Y.}\ \bibnamefont
  {Guslienko}}, \bibinfo {author} {\bibfnamefont {K.-S.}\ \bibnamefont {Lee}},
  \ and\ \bibinfo {author} {\bibfnamefont {S.-K.}\ \bibnamefont {Kim}},\ }\href
  {\doibase 10.1103/PhysRevLett.100.027203} {\bibfield  {journal} {\bibinfo
  {journal} {Phys. Rev. Lett.}\ }\textbf {\bibinfo {volume} {100}},\ \bibinfo
  {pages} {027203} (\bibinfo {year} {2008})}\BibitemShut {NoStop}%
\bibitem [{\citenamefont {Thiele}(1973)}]{Thiele1973}%
  \BibitemOpen
  \bibfield  {author} {\bibinfo {author} {\bibfnamefont {A.}~\bibnamefont
  {Thiele}},\ }\href {\doibase 10.1103/physrevlett.30.230} {\bibfield
  {journal} {\bibinfo  {journal} {Phys. Rev. Lett.}\ }\textbf {\bibinfo
  {volume} {30}},\ \bibinfo {pages} {230￢ﾀﾓ233} (\bibinfo {year}
  {1973})}\BibitemShut {NoStop}%
\bibitem [{\citenamefont {Huber}(1982)}]{Huber1982}%
  \BibitemOpen
  \bibfield  {author} {\bibinfo {author} {\bibfnamefont {D.~L.}\ \bibnamefont
  {Huber}},\ }\href@noop {} {\bibfield  {journal} {\bibinfo  {journal} {Phys.
  Rev. B}\ }\textbf {\bibinfo {volume} {26}},\ \bibinfo {pages} {3758}
  (\bibinfo {year} {1982})}\BibitemShut {NoStop}%
\bibitem [{\citenamefont {Dussaux}\ \emph {et~al.}(2012)\citenamefont
  {Dussaux}, \citenamefont {Khvalkovskiy}, \citenamefont {Bortolotti},
  \citenamefont {Grollier}, \citenamefont {Cros},\ and\ \citenamefont
  {Fert}}]{Dussaux2013}%
  \BibitemOpen
  \bibfield  {author} {\bibinfo {author} {\bibfnamefont {A.}~\bibnamefont
  {Dussaux}}, \bibinfo {author} {\bibfnamefont {A.~V.}\ \bibnamefont
  {Khvalkovskiy}}, \bibinfo {author} {\bibfnamefont {P.}~\bibnamefont
  {Bortolotti}}, \bibinfo {author} {\bibfnamefont {J.}~\bibnamefont
  {Grollier}}, \bibinfo {author} {\bibfnamefont {V.}~\bibnamefont {Cros}}, \
  and\ \bibinfo {author} {\bibfnamefont {A.}~\bibnamefont {Fert}},\ }\href
  {\doibase 10.1103/PhysRevB.86.014402} {\bibfield  {journal} {\bibinfo
  {journal} {Phys. Rev. B}\ }\textbf {\bibinfo {volume} {86}},\ \bibinfo
  {pages} {014402} (\bibinfo {year} {2012})}\BibitemShut {NoStop}%
\bibitem [{\citenamefont {Sukhostavets}\ \emph {et~al.}(2013)\citenamefont
  {Sukhostavets}, \citenamefont {Pigeau}, \citenamefont {Sangiao},
  \citenamefont {de~Loubens}, \citenamefont {Naletov}, \citenamefont {Klein},
  \citenamefont {Mitsuzuka}, \citenamefont {Andrieu}, \citenamefont
  {Montaigne},\ and\ \citenamefont {Guslienko}}]{Sukhostavets2013}%
  \BibitemOpen
  \bibfield  {author} {\bibinfo {author} {\bibfnamefont {O.~V.}\ \bibnamefont
  {Sukhostavets}}, \bibinfo {author} {\bibfnamefont {B.}~\bibnamefont
  {Pigeau}}, \bibinfo {author} {\bibfnamefont {S.}~\bibnamefont {Sangiao}},
  \bibinfo {author} {\bibfnamefont {G.}~\bibnamefont {de~Loubens}}, \bibinfo
  {author} {\bibfnamefont {V.~V.}\ \bibnamefont {Naletov}}, \bibinfo {author}
  {\bibfnamefont {O.}~\bibnamefont {Klein}}, \bibinfo {author} {\bibfnamefont
  {K.}~\bibnamefont {Mitsuzuka}}, \bibinfo {author} {\bibfnamefont
  {S.}~\bibnamefont {Andrieu}}, \bibinfo {author} {\bibfnamefont
  {F.}~\bibnamefont {Montaigne}}, \ and\ \bibinfo {author} {\bibfnamefont
  {K.~Y.}\ \bibnamefont {Guslienko}},\ }\href
  {http://dx.doi.org/10.1103/PhysRevLett.111.247601} {\bibfield  {journal}
  {\bibinfo  {journal} {Phys. Rev. Lett.}\ }\textbf {\bibinfo {volume} {111}},\
  \bibinfo {pages} {247601} (\bibinfo {year} {2013})}\BibitemShut {NoStop}%
\bibitem [{\citenamefont {Pigeau}(2012)}]{PigeauThesis}%
  \BibitemOpen
  \bibfield  {author} {\bibinfo {author} {\bibfnamefont {B.}~\bibnamefont
  {Pigeau}},\ }\emph {\bibinfo {title} {Magnetic vortex dynamics in
  nanostructures}},\ \href@noop {} {Ph.D. thesis},\ \bibinfo  {school}
  {Universite de Paris-Sud} (\bibinfo {year} {2012})\BibitemShut {NoStop}%
\bibitem [{\citenamefont {Buchanan}\ \emph {et~al.}(2007)\citenamefont
  {Buchanan}, \citenamefont {Grimsditch}, \citenamefont {Fradin}, \citenamefont
  {Bader},\ and\ \citenamefont {Novosad}}]{Buchanan2007}%
  \BibitemOpen
  \bibfield  {author} {\bibinfo {author} {\bibfnamefont {K.~S.}\ \bibnamefont
  {Buchanan}}, \bibinfo {author} {\bibfnamefont {M.}~\bibnamefont
  {Grimsditch}}, \bibinfo {author} {\bibfnamefont {F.~Y.}\ \bibnamefont
  {Fradin}}, \bibinfo {author} {\bibfnamefont {S.~D.}\ \bibnamefont {Bader}}, \
  and\ \bibinfo {author} {\bibfnamefont {V.}~\bibnamefont {Novosad}},\ }\href
  {\doibase 10.1103/PhysRevLett.99.267201} {\bibfield  {journal} {\bibinfo
  {journal} {Phys. Rev. Lett.}\ }\textbf {\bibinfo {volume} {99}},\ \bibinfo
  {pages} {267201} (\bibinfo {year} {2007})}\BibitemShut {NoStop}%
\bibitem [{\citenamefont {Drews}\ \emph {et~al.}(2012)\citenamefont {Drews},
  \citenamefont {Kr{\"u}ger}, \citenamefont {Selke}, \citenamefont {Kamionka},
  \citenamefont {Vogel}, \citenamefont {Martens}, \citenamefont {Merkt},
  \citenamefont {MÃ¶ller},\ and\ \citenamefont {Meier}}]{Drews2012}%
  \BibitemOpen
  \bibfield  {author} {\bibinfo {author} {\bibfnamefont {A.}~\bibnamefont
  {Drews}}, \bibinfo {author} {\bibfnamefont {B.}~\bibnamefont {Kr{\"u}ger}},
  \bibinfo {author} {\bibfnamefont {G.}~\bibnamefont {Selke}}, \bibinfo
  {author} {\bibfnamefont {T.}~\bibnamefont {Kamionka}}, \bibinfo {author}
  {\bibfnamefont {A.}~\bibnamefont {Vogel}}, \bibinfo {author} {\bibfnamefont
  {M.}~\bibnamefont {Martens}}, \bibinfo {author} {\bibfnamefont
  {U.}~\bibnamefont {Merkt}}, \bibinfo {author} {\bibfnamefont
  {D.}~\bibnamefont {MÃ¶ller}}, \ and\ \bibinfo {author} {\bibfnamefont
  {G.}~\bibnamefont {Meier}},\ }\href@noop {} {\bibfield  {journal} {\bibinfo
  {journal} {Phys. Rev. B}\ }\textbf {\bibinfo {volume} {85}},\ \bibinfo
  {pages} {144417} (\bibinfo {year} {2012})}\BibitemShut {NoStop}%
\bibitem [{\citenamefont {Guslienko}\ \emph {et~al.}(2010)\citenamefont
  {Guslienko}, \citenamefont {Heredero},\ and\ \citenamefont
  {Chubykalo-Fesenko}}]{Guslienko2010}%
  \BibitemOpen
  \bibfield  {author} {\bibinfo {author} {\bibfnamefont {K.~Y.}\ \bibnamefont
  {Guslienko}}, \bibinfo {author} {\bibfnamefont {R.~H.}\ \bibnamefont
  {Heredero}}, \ and\ \bibinfo {author} {\bibfnamefont {O.}~\bibnamefont
  {Chubykalo-Fesenko}},\ }\href@noop {} {\bibfield  {journal} {\bibinfo
  {journal} {Phys. Rev. B}\ }\textbf {\bibinfo {volume} {82}},\ \bibinfo
  {pages} {014402} (\bibinfo {year} {2010})}\BibitemShut {NoStop}%
\end{thebibliography}
\end{document}